\documentclass[pra,twocolumn,showpacs,preprintnumbers,amsmath,amssymb,floatfix,superscriptaddress] {revtex4-1}

\usepackage{graphicx}
\usepackage{dcolumn}
\usepackage{bm,bbm}
\usepackage{color}

\begin{document}


\title{Long range propagation of ultrafast, ionizing laser pulses in a resonant nonlinear medium}%

\author{G. Demeter}
\email{demeter.gabor@wigner.hu}
\affiliation{Wigner Research Centre for Physics, Budapest, Hungary}

\author{J. T. Moody}
\affiliation{Max Planck Institute for Physics, Munich, Germany}

\author{M. \'{A}. Kedves}
\affiliation{Wigner Research Centre for Physics, Budapest, Hungary}

\author{B. R\'{a}czkevi}
\affiliation{Wigner Research Centre for Physics, Budapest, Hungary}

\author{M. Aladi}
\affiliation{Wigner Research Centre for Physics, Budapest, Hungary}

\author{A.-M. Bachmann}
\affiliation{Max Planck Institute for Physics, Munich, Germany}

\author{F. Batsch}
\affiliation{Max Planck Institute for Physics, Munich, Germany}

\author{F. Braunm\"{u}ller}
\affiliation{Max Planck Institute for Physics, Munich, Germany}

\author{G. P. Djotyan}
\affiliation{Wigner Research Centre for Physics, Budapest, Hungary}

\author{V. Fedosseev}
\affiliation{CERN, Geneva, Switzerland}

\author{F. Friebel}
\affiliation{CERN, Geneva, Switzerland}

\author{S. Gessner}
\affiliation{CERN, Geneva, Switzerland}
\affiliation{SLAC National Accelerator Laboratory, Menlo Park, California, USA}

\author{E. Granados}
\affiliation{CERN, Geneva, Switzerland}

\author{E. Guran}

\author{M. H\"{u}ther}
\affiliation{Max Planck Institute for Physics, Munich, Germany}

\author{V. Lee}
\affiliation{University of Colorado Boulder, Colorado, USA}

\author{M. Martyanov}
\affiliation{Max Planck Institute for Physics, Munich, Germany}
\affiliation{CERN, Geneva, Switzerland}

\author{P. Muggli}
\affiliation{Max Planck Institute for Physics, Munich, Germany}

\author{E. \"{O}z}
\affiliation{Max Planck Institute for Physics, Munich, Germany}

\author{H. Panuganti}
\affiliation{CERN, Geneva, Switzerland}

\author{L. Verra}
\affiliation{Max Planck Institute for Physics, Munich, Germany}
\affiliation{CERN, Geneva, Switzerland}
\affiliation{Technical University Munich, Munich, Germany}

\author{G. Zevi Della Porta}
\affiliation{CERN, Geneva, Switzerland}
 

\date{\today}

\begin{abstract}
We study the propagation of 0.05-1 TW power, ultrafast laser pulses in a 10 meter long rubidium vapor cell. The central wavelength of the laser is resonant with the $D_2$ line of rubidium and the peak intensity is in the $10^{12}-10^{14} \mathrm{~W/cm^2}$ range, enough to create a plasma channel with single electron ionization. We observe the absorption of the laser pulse for low energy, a regime of transverse confinement of the laser beam by the strong resonant nonlinearity for higher energies and the transverse broadening of the output beam when the {resonant nonlinearity ceases due to the valence electrons being all removed during ionization}. We compare experimental observations of transmitted pulse energy and transverse fluence profile with the results of computer simulations modeling pulse propagation. 
We find a qualitative agreement between theory and experiment that corroborates the validity of our propagation model.
While the quantitative differences are substantial, the results show that the model can be used to interpret the observed phenomena in terms of self-focusing and channeling of the laser pulses by the {saturable, resonant nonlinearity.}  
\end{abstract}

\maketitle

\section{Introduction}
\label{intro}

Particle acceleration in plasma wakefields is a concept about four decades old, that is flourishing today in diverse directions. The intense work going on in a multitude of places worldwide is fueled by a series of scientific and technical advances that hold the promise to transfer the plasma wakefield accelerator scheme to use in applications for science and technology in the near future. Prospective applications for the scheme range from compact, high-quality particle beam sources for high-energy physics to x-ray light sources such as Compton scattering and free electron lasers \cite{Albert2021}. Large scale international collaborations labor to turn promise into reality \cite{awake,eupraxia,Assmann2020}.

One experimental concept aimed at high-energy physics, the Advanced Proton Driven Wakefield Acceleration Experiment (AWAKE) at CERN is the first wakefield accelerator to use a high-energy proton beam driver to accelerate an electron bunch \cite{Caldwell2009,Caldwell2016, Gschwendtner2016}. The plasma in this device serves two purposes: it first modulates the long proton driver to generate a sequence of microbunches via seeded self modulation and second serves as the energy exchange medium where the microbunches drive wakefields that can accelerate the electrons. Run 1 of the AWAKE experiment used a single, 10 meter long plasma chamber to fulfill both these purposes \cite{Adli2018}, while the Run 2 phase of AWAKE will eventually use two separate 10 meter long plasmas, a `modulator' and an `accelerator' \cite{Muggli2020}. Creating a plasma channel of this length with the precisely engineered density distribution required is very difficult. The technology currently utilized at AWAKE involves creating rubidium vapor with the prescribed density distribution and ionizing it with a high-intensity,  ultra-short laser pulse. Rubidium has a single outer electron that is easily removed ($E_{1}=4.18 \mathrm{~eV}$) and a closed shell underneath difficult to break ($E_{2}=27.29 \mathrm{~eV}$ for the second electron), so single electron ionization of nearly all of the atoms in a volume is expected \cite{Muggli2017}. Initially engineered vapor density then translates into precisely defined plasma density. 

However, creating meter scale, optical-field-ionized plasmas for wakefield acceleration is challenging as
the propagation of high power laser pulses in gaseous media is rich in complex phenomena. The strong nonlinear interaction that arises leads, among others, to filamentation: the confinement of laser energy along thin, self-guided structures \cite{Berge1998, Berge2007, Couairon2007, Kandidov2009, Kolesik2013}.  The archetypal scenario for filamentation is the dynamical competition between a focusing Kerr nonlinearity, diffraction and defocusing processes (e.g. plasma defocusing or some higher order defocusing nonlinearity) or intensity clamping processes (e.g. ionization losses).   
In practice the picture is usually complex, there are many possibilities in different media as to what processes define or contribute to laser filamentation and this field is still a lively area today both theoretically and experimentally. 
In addition, plasma dynamical phenomena are sometimes called upon to help guide the ionizing pulses along the prescribed axis to obtain a plasma channel that fulfills wakefield acceleration requirements \cite{Picksley2020}.

The laser pulse propagation scenario considered here is peculiar and highly interesting because the TW class Ti:Sa laser system of the AWAKE facility has a central wavelength of $780\mathrm{~nm}$ \cite{Muggli2017}, coinciding with the rubidium $D_2$ line, the strong dipole transition between the ground state and the $5\mathrm{P}_{3/2}$  {state, the first excited state. Transition frequencies from $5\mathrm{P}_{3/2}$ to higher lying bound states are also within the laser bandwidth. These single-photon resonances make the nonlinear material response of neutral atoms much stronger compared to the nonresonant case, but as the valence electron is removed due to ionization, resonant interaction ceases so the nonlinearity is, in effect, {\em saturable}.} This situation has not been studied in depth in the context of laser filamentation. Filamentation in the presence of multiphoton resonances has been studied recently \cite{Doussot2016, Doussot2017}, demonstrating the highly nontrivial effects of these resonances on the physics of pulse propagation. 
But a single photon resonance from the ground state is very different as it provides absorption and strong optical nonlinearity even at low intensity. This is more the realm of traditional resonant nonlinear optics \cite{Boshier1982, Lamb1971, Lamare1994, Delagnes2008}, which has also been extensively studied, but for much smaller light intensities (without ionization) and longer pulse lengths. In contrast to the non-resonant case, where the medium is effectively transparent until light intensity is high enough to ionize, the resonant medium is absorbing  {even at low intensities, but is rendered effectively transparent, when all atoms have shed their valence electrons. } 
The traditional filamentation scenario results in the weak ionization of a domain much narrower than the laser beam diameter, diffraction and plasma gradient defocusing both playing a considerable role in determining the plasma channel radius. The present scenario with single photon resonances on the other hand leads to single-electron ionization of all atoms in a channel on the same transverse scale as the laser beam, plasma gradient and diffraction playing a less significant role. Overall, the result is much more favorable for wakefield acceleration.

A theoretical model has been developed recently to describe this scenario and it was used to study numerically plasma channel formation in rubidium vapor for large propagation distances \cite{Demeter2019}. Self-focusing at low intensity,  {self-channeling due to the transparency of ionized vapor at higher intensities} and interesting quasiperiodic oscillations of the plasma channel radius were predicted. Should the model eventually prove accurate enough to have quantitative predictive power, the scalability and limits of plasma channel creation using high intensity, resonant laser pulses could be evaluated for the benefit of the plasma wakefield acceleration community.

In this paper we present an experimental study of resonant, TW scale power laser pulse propagation in a 10 meter long rubidium vapor performed at the AWAKE facility at CERN. Observations were made for several vapor density values and using a detailed scan of input pulse energy. Several distinct interaction regimes were identified in the experimental results, the first that resulted from an almost complete absorption of a weak pulse, one that resulted from a complete saturation of the medium for a large energy pulse and two intermediate regimes. Computer simulations were performed with matching parameters using a theory almost identical to that presented in \cite{Demeter2019}. We contrast experimental observations with numerical results and discuss similarities and discrepancies between theory and experiment.

\section{Experiment}

\subsection{Setup}

\begin{figure}[htb]
\includegraphics[width=0.47\textwidth]{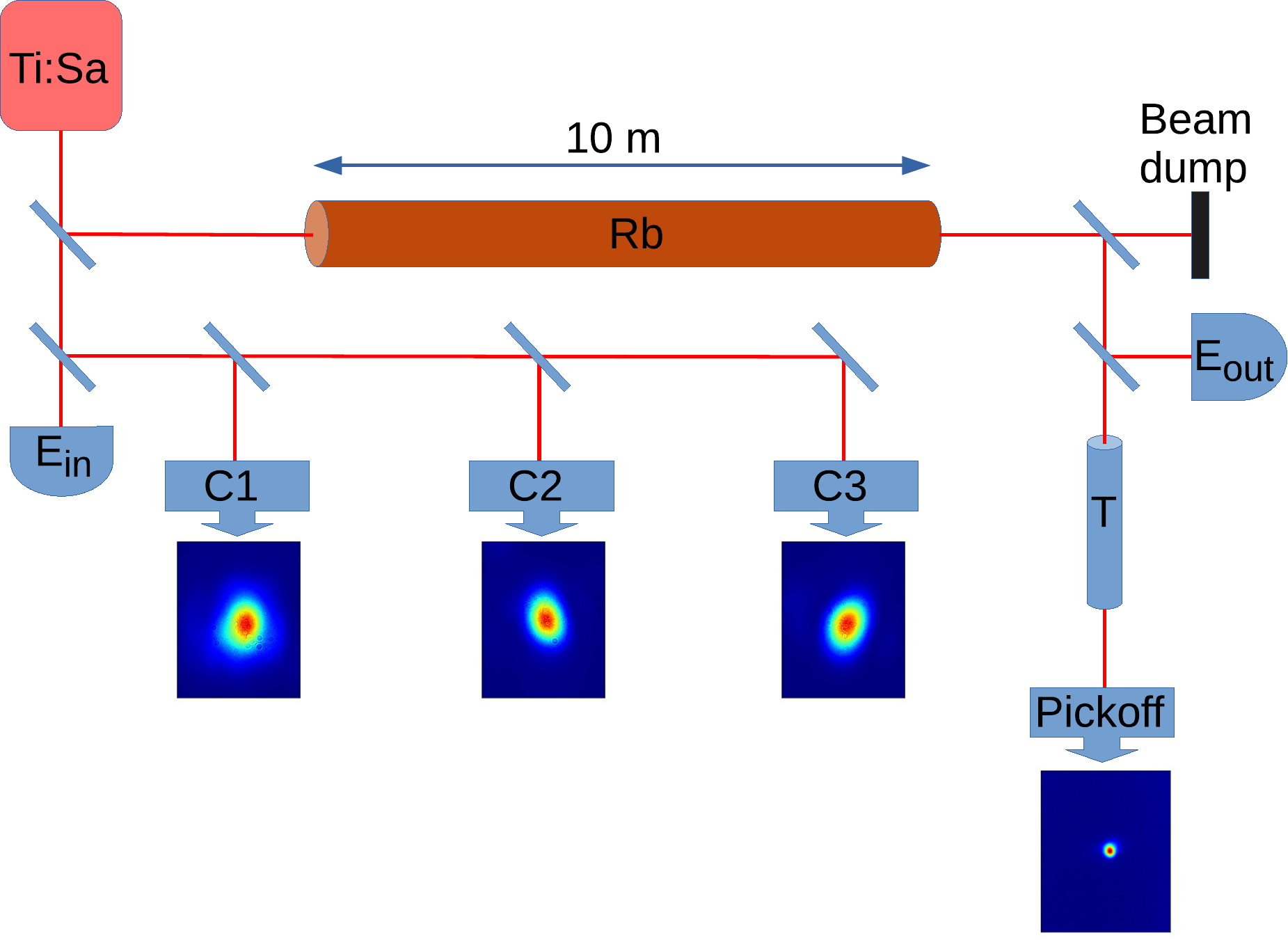}
\caption{Sketch of the experimental setup. $E_{in}$ and $E_{out}$ are the input and output energy meters, respectively. Cameras in the virtual laser line correspond to the vapor chamber entrance (C1), center (C2) and exit (C3). Transmitted light distribution is detected by the pickoff camera after the imaging telescope (T). Images below the cameras are illustrative images from a single pulse measurement. Optical paths are not drawn to scale, the optical path from the final focusing telescope of the Ti:Sa laser to the vapor source entrance (and the C1 camera) is $\approx$40 m.}
\label{setup}       
\end{figure}

Experiments were performed using components of AWAKE Run 1 \cite{Gschwendtner2016, Muggli2017}, when the proton and electron beams were not in operation. Pulses from a Ti:Sa laser system with 780 nm central wavelength and 120 fs pulse duration were focused by a mismatched telescope into a 10 m long rubidium vapor source, through a 10 mm diameter aperture. The beam waist was approximately $w_0=1.5$ mm, waist location at around $z_0=7$ m from the upstream end of the vapor source (slightly variable location). The temperature controlled rubidium reservoirs and walls of the source provided a highly homogeneous vapor, rubidium density was regulated by setting the temperature of the reservoirs and measured using white-light interferometry \cite{Oz2014, Plyushchev2017, Batsch2018}.
Laser pulse energy in the experiment was regulated from 0 mJ to 120 mJ by a waveplate and two Brewster polarizers between the last amplifier and the compressor. Transmission from one of the mirrors in the laser line upstream of the vapor chamber was used to set up a virtual laser line with an energy meter and three cameras to record the transverse laser distribution at propagation distances corresponding precisely to the entrance, center and exit of the vapor source.  These were used to collect images of the `virtual entrance', `virtual center' and `virtual exit' of the vapor source, recording the transverse distribution of the beam as it would be seen propagating in vacuum across the chamber. (C1, C2 and C3 on Fig. \ref{setup} respectively. Cameras were Basler acA1920-40gm, image resolution determined by the pixel size 5.86 $\mu$m, due to direct beam input.) The input energy meter ($E_{in}$) was calibrated by placing a direct energy meter into the laser line when the vacuum system was open. Ten meters downstream of the end of the vapor chamber, the front surface of a pickoff wedge placed into the beamline before the beam dump diverted $\sim.5\%$ of the laser pulse to the output energy meter ($E_{out}$) and to a two-lens imaging system. The lenses were used to create an image of the vapor source exit on the pickoff camera that recorded the transverse energy distribution of the pulse after propagating through the vapor, image resolution was about 40 $\mu$m. The reading on the output energy meter ($E_{out}$) was calibrated to the reading on the input one ($E_{in}$) by a series of measurements with the valves of the rubidium reservoirs attached to the chamber closed and the chamber at room temperature. We estimate that under these conditions the residual rubidium vapor absorbs at most about a $\mathrm{\mu J}$ of laser energy. 
Variable filters were used on the virtual laser line cameras and the pickoff camera to prevent saturation. Transverse energy distributions on the virtual laser line cameras were scaled to physical units using the known camera pixel size. Images on the pickoff camera were scaled using a scaling factor derived by comparing the virtual exit images (C3) to the corresponding pickoff images for measurements that were performed with residual rubidium vapor. The vapor has a negligible influence on the laser beam profile in this case.  {More details on the calibration process and a more accurate drawing of the experimental setup can be found in the Supplemental Material, which includes Ref. \cite{Alcock1984}.}

\subsection{Measurements and observations}

The properties of the laser pulses were measured after propagating along the vapor source as a function of $E_{in}$ at three different values of vapor density $\mathcal{N}=1.87\cdot 10^{14}\mathrm{~cm}^{-3}$, $\mathcal{N}=4.895\cdot 10^{14}\mathrm{~cm}^{-3}$, and $\mathcal{N}=6.6\cdot 10^{14}\mathrm{~cm}^{-3}$ - these values correspond to the ones used in the wakefield experiments. The transverse energy distributions (fluence profiles) at the three cameras of the virtual laser line (C1, C2 and C3) and that of the transmitted pulse (pickoff camera) were recorded, along with the corresponding values of $E_{in}$ and $E_{out}$. Width parameters to characterize the overall transverse size of the fluence profiles $\mathcal{F}(x,y)$ were then calculated for each image by function fits to the measured distributions. The nonlinear least-squares problem was solved by a Trust Region Reflective algorithm contained in the \texttt{scipy.optimize} package, implemented in Python  \cite{SciPy2020}. For comparison with the numerical calculations, an axisymmetric Gaussian distribution  $\mathcal{G} = A_0\exp(-2((x-x_0)^2+(y-y_0)^2)/\sigma^2) + C$ was used in the fit to approximate $\mathcal{F}(x,y)$ and obtain a single $\sigma$ width parameter. Peak fluence $\mathcal{F}_{max}$ was calculated from the maximum pixel count of the images after background deduction.

\begin{figure}[htb]
\includegraphics[width=0.48\textwidth]{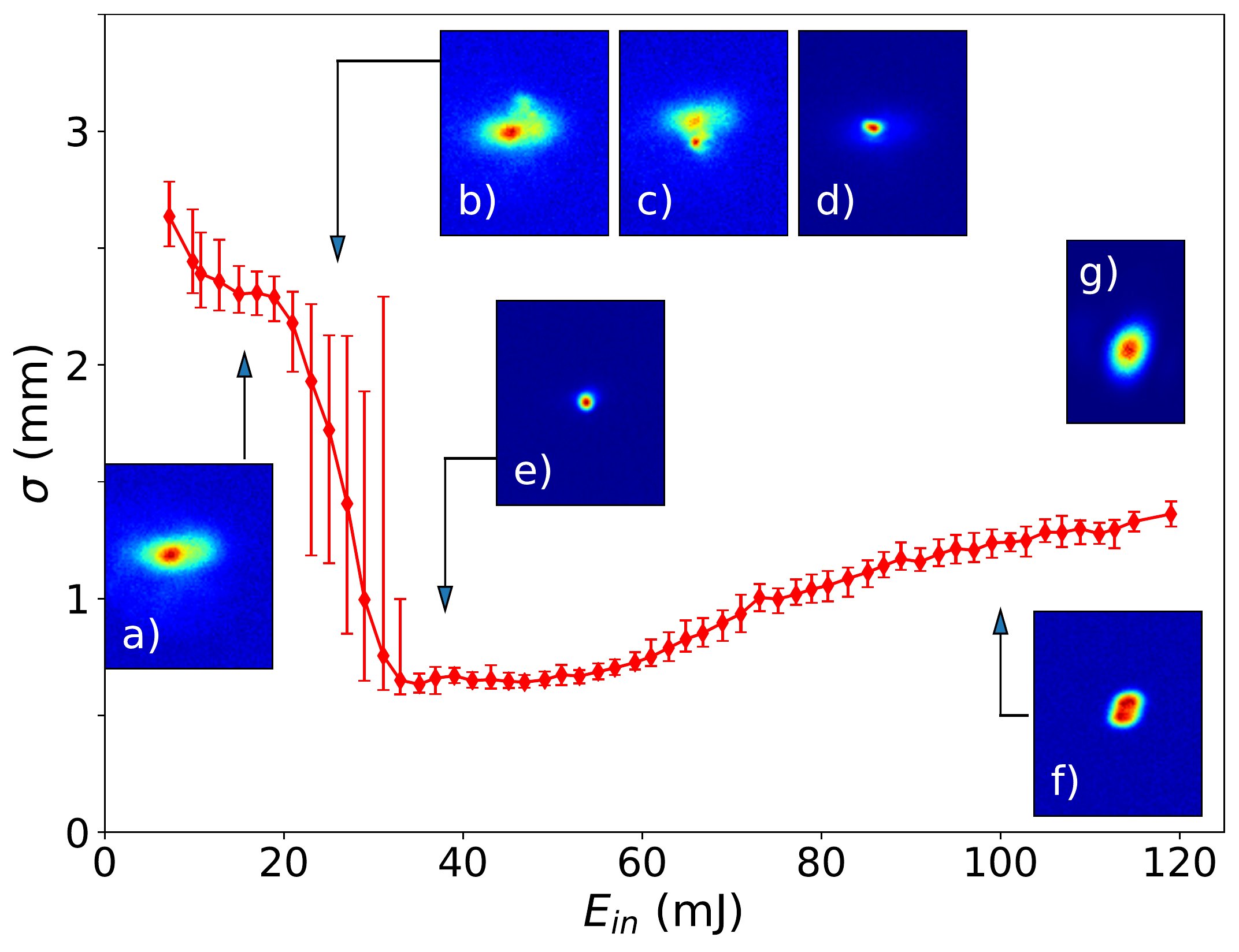}
\caption{Width parameter $\sigma$ from Gaussian fit for transmitted laser pulses, 
$\mathcal{N}=6.6\cdot 10^{14}\mathrm{cm}^{-3}$. Mean values of binned data shown with asymmetric standard deviation indicated. Insets depict pickoff camera images for selected shots with: a) $E_{in}=15.6\mathrm{~mJ}$, b)-d) $E_{in}=26\mathrm{~mJ}$, e) $E_{in}=38\mathrm{~mJ}$ and f) $E_{in}=100\mathrm{~mJ}$, each also marked by the arrows. Aspect ratio of the insets corresponds to the detector physical aspect ratio, color coding of individual images are unique, scaled to individual image maxima. Inset g) depicts the virtual exit camera image (C3) for $E_{in}=38\mathrm{~mJ}$, drawn to the same spatial scale.}
\label{pickoff_width}       
\end{figure}

An example of the information obtained after processing the data can be seen on Fig. \ref{pickoff_width}, created from $\mathcal{N}=6.6\cdot 10^{14}\mathrm{~cm}^{-3}$ vapor density shots. Values of $\sigma$ calculated for individual shots have been binned with respect to input energy and bin averages plotted with asymmetric error bars showing the standard deviation of data below and above the mean separately.
 {Fluctuations associated with the transition around $E_{in}=20-30$ mJ are very high and asymmetric around the mean, because they are associated with the random occurrence of narrow and wide transmitted beams with a changing relative frequency. }
Individual bins typically contain the data of 20-40 individual shots, with a few between 10-20 shots or 40-54 shots. The last three data points ($E_{in}>112\mathrm{~mJ}$) represent bins of 2-4 shots only. Insets depict camera images of the transmitted pulse for a few selected shots with arrows pointing to the region of input energy from where they were selected. They can be considered 'typical' images for the given region, that are representative of the transmitted laser beam transverse shapes. In addition, a single inset depicts the image recorded by the virtual exit camera (C3), drawn to the same spatial scale as insets depicting pickoff camera images, so laser pulse transverse size can be compared. 

\begin{figure}[htb]
\includegraphics[width=0.47\textwidth]{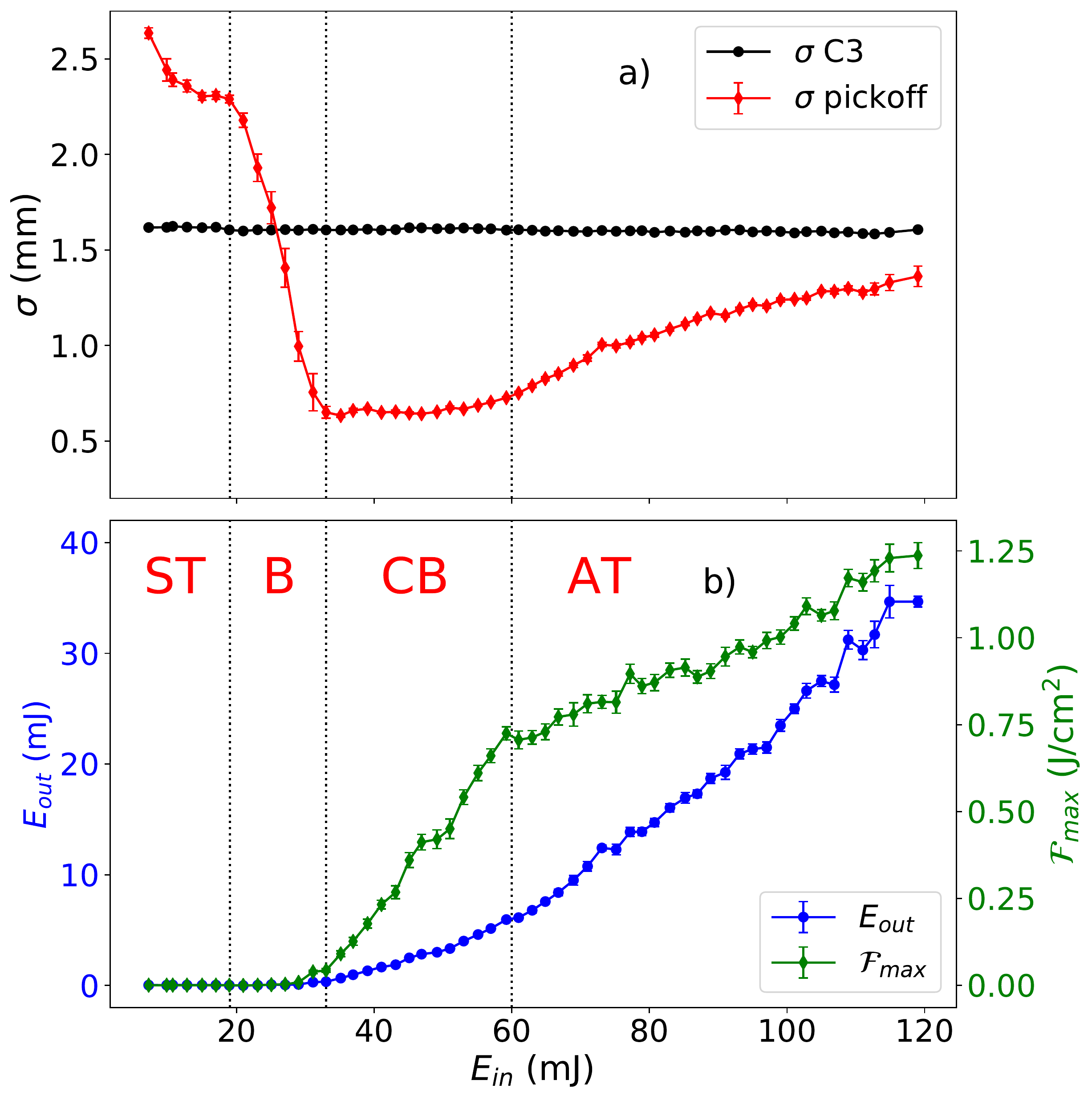}
\caption{a) Width parameter $\sigma$ for transmitted pulse and virtual exit camera image as a function of input pulse energy for $\mathcal{N}=6.6\cdot 10^{14}\mathrm{cm}^{-3}$ vapor density shots. b) Transmitted laser pulse energy  (left axis) and peak fluence (right axis). The points depict averages of binned data,  error bars mark the error of the mean. Vertical dotted lines mark the approximate domain boundaries, which are labeled as: `ST' for sub-threshold domain, `B' for breakthrough domain, `CB' for confined beam domain and `AT' for asymptotic transparency domain.}
\label{N7e14}       
\end{figure}

Figure \ref{N7e14} depicts a) the same transmitted pulse $\sigma$, together with the $\sigma$ parameter of the virtual exit camera for reference and b) the transmitted pulse $E_{out}$ and the peak fluence $\mathcal{F}_{max}$. 
The curves were created by binning the data of individual shots, markers show the bin mean and error bars correspond to the error of the mean. Several distinct regions are visible with respect to $E_{in}$, separated by dotted vertical lines drawn to guide the eye. For the lowest values of $E_{in}$, laser pulses are broadened in the transverse plane (see also insets a) and g) of Fig. \ref{pickoff_width}) with very low energy. In this region almost all of the energy is absorbed by the rubidium vapor, only frequency components sufficiently far from the resonance frequency of the $D_2$ transition may be transmitted. We will call this region the {\em sub-threshold domain}, labeled by `ST' on Fig. \ref{N7e14}. The next region shows a steep decrease of the average beam width, accompanied by large fluctuations,  the deviations from the mean are very asymmetric. This is caused by a `mixture' of output beam profiles, 
broad, low amplitude pulses may appear randomly as well as very sharp, narrow pulses as seen on 
Fig. \ref{pickoff_width}, insets b)-d). Narrow pulses appear only rarely initially and they appear more and more often as $E_{in}$ increases. Correspondingly, the probability that the transmitted pulse will be a broad, low amplitude one, decreases. Occasionally, traces of multiple sharp maxima appear on the transmitted pulse image as seen on Fig. \ref{pickoff_width}, inset c). We will call this region the {\em breakthrough domain}, labeled by `B' on Fig. \ref{N7e14}, which also shows that the sub-threshold and breakthrough domains are characterized by practically zero $E_{out}$ and $\mathcal{F}_{max}$. 

Above the breakthrough domain, for a substantial interval of $E_{in}$ the transmitted pulse $\sigma$ does not significantly increase, but $\mathcal{F}_{max}$ grows sharply and $E_{out}$ also starts to increase. The transmitted beam shape is also much more axisymmetric (inset e) of Fig. \ref{pickoff_width}) than the somewhat elongated, elliptical wide beams in the sub-threshold domain. We will call this region the {\em confined beam domain}, labeled by `CB' on Fig. \ref{N7e14}. Finally, above this domain the output beam starts to broaden again (inset f) of Fig. \ref{pickoff_width}), $E_{out}$ starts increasing substantially and the rate at which $\mathcal{F}_{max}$ grows decreases (Fig. \ref{N7e14} b) ). The transmitted beam width converges slowly to the original beam width observed on the virtual exit camera, suggesting that as the medium nonlinearity is saturated by complete conversion to Rb$^{1+}$ ions, the effect on the propagating pulse becomes less and less (Fig. \ref{N7e14} a) ). We will call this region the {\em asymptotic transparency domain}, labeled by `AT' on Fig. \ref{N7e14}.

\begin{figure}[htb]
\includegraphics[width=0.47\textwidth]{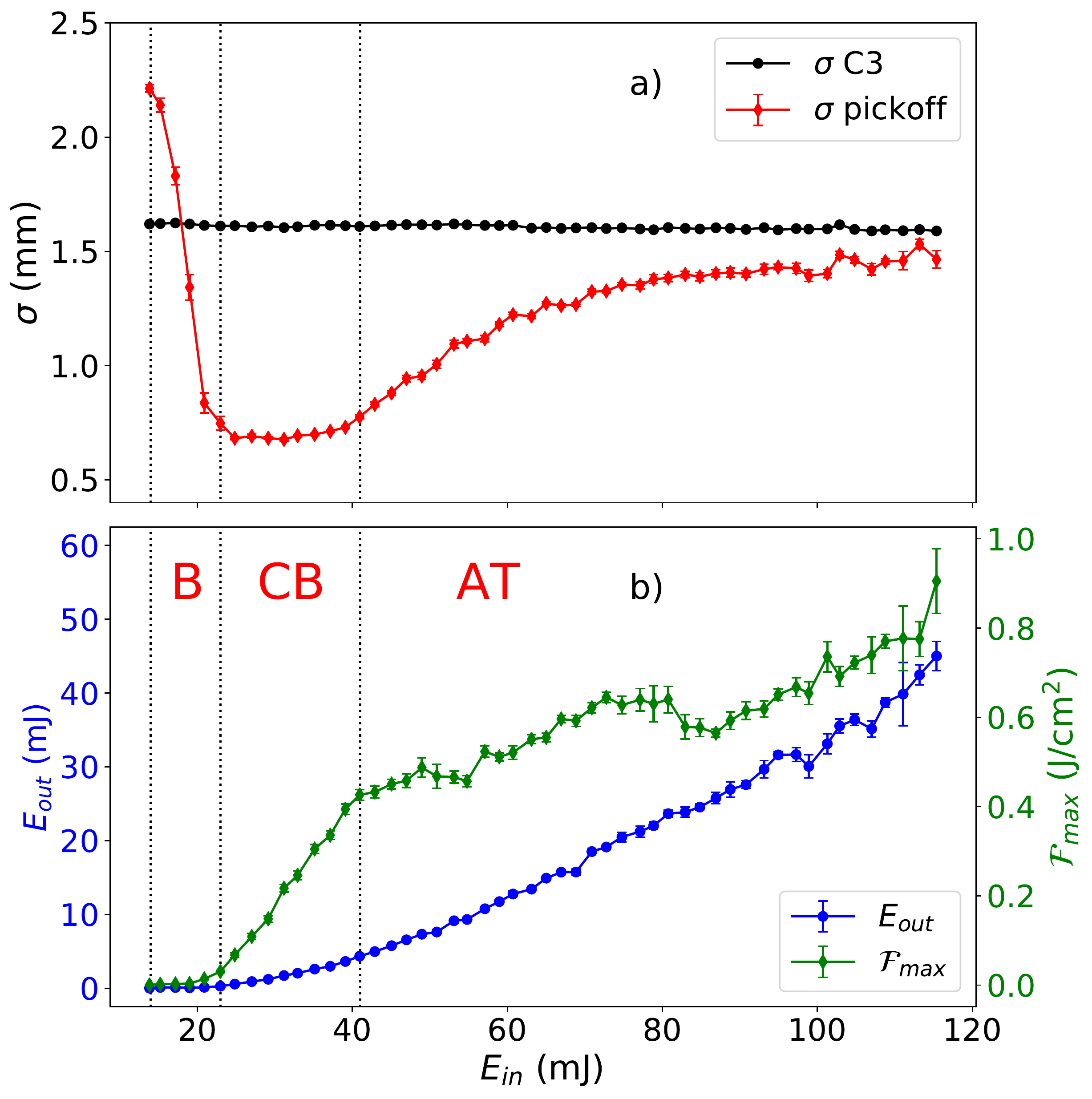}
\caption{Plot identical to Fig. \ref{N7e14} for $\mathcal{N}=4.895\cdot 10^{14}\mathrm{cm}^{-3}$ vapor density shots. The sub-threshold domain is not captured by the dataset. }
\label{N5e14}       
\end{figure}

Figure \ref{N5e14} depicts the same plots for $\mathcal{N} = 4.895\cdot 10^{14}\mathrm{~cm^{-3}}$ vapor density. The region of the confined beam domain is shorter here and evidently the sub-threshold domain is not captured by the data set. Convergence to the original beam width is faster for large energies. The minimum transmitted beam width observed (at the start of the confined beam domain) is $\sigma=0.633\pm0.009 \mathrm{~mm}$ for $\mathcal{N}=6.6\cdot 10^{14}\mathrm{cm}^{-3}$ vapor density and $\sigma=0.677\pm0.007 \mathrm{~mm}$ for $\mathcal{N}=4.895\cdot 10^{14}\mathrm{cm}^{-3}$ vapor density. For the lowest vapor density measurements $\mathcal{N}=1.87\cdot 10^{14}\mathrm{cm}^{-3}$ 
the systematic changes described above are not captured by the dataset, but instead there is a rapid early transition to the asymptotic transparency regime (see Fig. \ref{sigmacurves} a) ).

\section{Theoretical framework}

A theory for calculating the long-range propagation of ultrashort, ionizing laser pulses in rubidium vapor under the specific condition when the laser frequency is resonant with an atomic transition from the ground state 
has recently been developed \cite{Demeter2019}. This theory is substantially different from the approach usually used for calculating the propagation of intense laser pulses in atmospheric gases where ionization and laser pulse filamentation can be observed. In this case, laser pulses are intense enough to ionize via multiphoton or tunnel 
ionization directly from the ground state ($I_{max}\gtrsim\mathrm{~TW/cm^2}$), but the atomic response has a major 
contribution from Rabi-oscillation type transitions on single photon resonances.
Here we present only a very concise account of the theory we use, as it is almost the same as the one presented in \cite{Demeter2019} in greater detail.

We consider the propagation along the $z$ direction of a linearly polarized laser pulse in the paraxial approximation, 
assuming axial symmetry - we denote the single transverse coordinate with $r$. We separate the central frequency of the laser $\omega_0 = k_0 c$ from the electric field in the form   
$E(r,z,t)=\frac{1}{2}\mathcal{E}(r,z,t)\exp(ik_0z-\omega_0t)+c.c.$ ($\mathcal{E}(r,z,t)$ is a complex envelope function) and do the same for medium polarization terms $\mathcal{P}(r,z,t)$,  
$\mathcal{R}(r,z,t)$ and  $\mathcal{Q}(r,z,t)$ to be defined later. Transforming from $(r,z,t)$ to a new reference frame $(r,\xi,\tau)$ with $\xi=z$ and $\tau=t-z/c$, we write the propagation equation for the time Fourier transform of the complex envelope function 
$\tilde{\mathcal{E}}(r,\xi,\omega)=\mathfrak{F}\{\mathcal{E}(r,\xi,\tau)\}$
(where $\mathfrak{F}\{\ldotp\} $ denotes the time-Fourier transform). 
We employ the Slowly Evolving Wave Approximation (SEWA) \cite{Brabec1997,Couairon2011} that allows the treatment of ultrashort pulses and sharp leading edges that may develop to arrive at the propagation equation:  
\begin{equation}
\begin{aligned}
\partial_\xi \tilde{\mathcal{E}}  = & 
\frac{i}{2k} \nabla_\perp^2\tilde{\mathcal{E}}
+ i\frac{k}{2\epsilon_0} \tilde{\mathcal{P}} \\
& -\eta_0\hbar\omega_0\mathcal{N}\tilde{\mathcal{Q}} 
- \frac{ik}{2}\frac{e^2\mathcal{N}}{\epsilon_0 m_e (\omega_0+\omega)^2} \tilde{\mathcal{R}}
\end{aligned}
\label{waveeq}
\end{equation}
Here $e,m_e$ are the elementary charge and electron mass,
$\epsilon_0, \eta_0$ the vacuum permittivity and impedance and $k = (\omega_0+\omega)/c$ is the wavenumber. The first term on the right-handside of Eq. \ref{waveeq} is due to diffraction, while the other three are due to the medium as detailed below.

\begin{figure}[htb]
\includegraphics[width=0.5\textwidth]{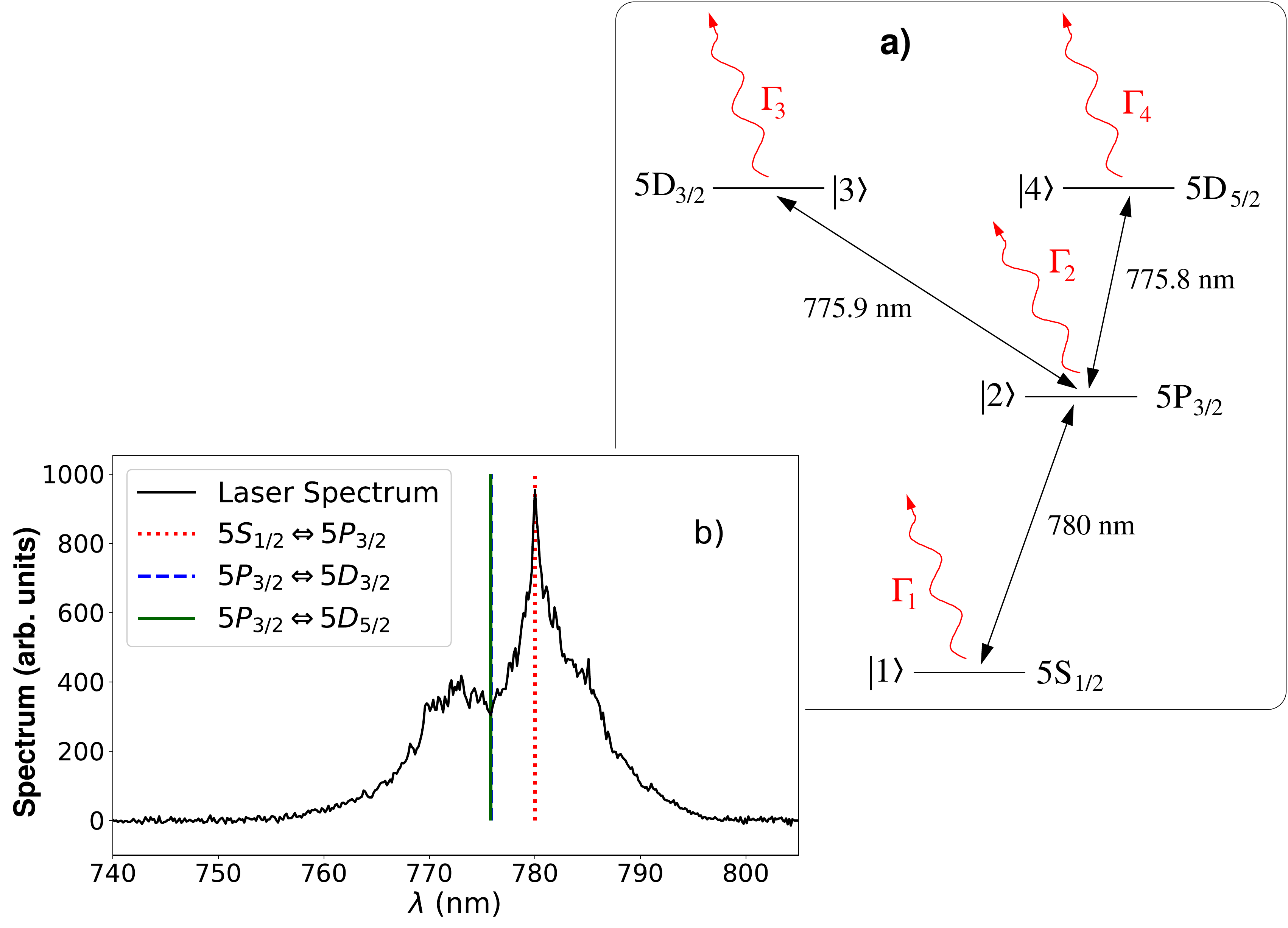}
\caption{a) Electronic levels of the rubidium atom that are included in the model and their numbering. Three 
excited states are resonantly accessible from the ground state, ionization leads to level loss from 
each of the levels. b) Measured spectrum of the ionizing laser oscillator before amplification and interaction with the vapor.  The resonance wavelengths included in the model are marked, lines $5P_{3/2} \Leftrightarrow 5D_{3/2}$ and $5P_{3/2} \Leftrightarrow 5D_{5/2}$ are a closely spaced doublet, difficult to resolve on this scale.}
\label{fig_Rblevels}       
\end{figure}

 Because a power law expansion of the medium polarization in terms of the field amplitude does not converge at resonance \cite{BoydNonlinOptics}, an explicit calculation of the atomic states' time dependence due to the applied field must be performed in order to obtain the transient response to the applied field. (The classical formula for anomalous dispersion in the vicinity of a resonance is valid only when the relevant timescales are larger than relaxation times.) 
 To this end, we employ a simplified atomic model that takes into account the resonant atomic transitions as well as multiphoton or tunnel ionization.
The model uses the ground state and the three excited states that are accessible from the ground state
via resonant transitions with wavelengths within the bandwidth of the laser light, denoted by $|j\rangle, j\in\{1,4\}$, shown in Fig. \ref{fig_Rblevels}. 
We define the atomic state $|\psi\rangle$ using probability amplitudes on the $|j\rangle$ basis with some convenient phases as:
\begin{equation}
\begin{aligned}
 |\psi(t)\rangle = & a_1(t)e^{-i(\omega_2-\omega_0)t}|1\rangle + a_2(t)e^{-i\omega_2t}|2\rangle \\
 & a_3(t)e^{-i(\omega_2+\omega_0)t}|3\rangle + a_4(t)e^{-i(\omega_2+\omega_0)t}|4\rangle
\end{aligned}
\end{equation}
where $\hbar\omega_2$ is the energy difference between the $\mathrm{5S_{1/2}}$ ground state and the $\mathrm{5P_{3/2}}$ first excited state. Using this notation,
the time evolution of the atomic state at any point in space is given by:
\begin{equation}
\begin{aligned}
 \partial_\tau a_1 =& -i\Delta_{21}a_1+\frac{i}{2\hbar}\mathcal{E}^*d_{21}a_2 - 
\frac{\Gamma_1}{2}a_1\\
 \partial_\tau a_2 =& \frac{i}{2\hbar}\bigl(\mathcal{E} d_{21}a_1 + 
   \mathcal{E}^*d_{32}a_3 + \mathcal{E}^*d_{42}a_4\bigr) - \frac{\Gamma_2}{2}a_2 \\
  \partial_\tau a_3 =& i\Delta_{32}a_3+\frac{i}{2\hbar}\mathcal{E} d_{32}a_2 - 
\frac{\Gamma_3}{2}a_3\\
 \partial_\tau a_4 =& i\Delta_{42}a_4+\frac{i}{2\hbar}\mathcal{E} d_{42}a_2 - \frac{\Gamma_4}{2}a_4
\end{aligned}
\label{schrodinger}
\end{equation}
Here the transition matrix elements $d_{kl}$ between atomic states and the frequency detunings from resonance frequencies $\Delta_{jk} = \omega_0-\omega_{jk}$ are material parameters obtained from the literature \cite{steckRb85, NIST, Safronova2004}. Their numerical values are collected in the appendix of \cite{Demeter2019}. The (intensity dependent) multiphoton ionization rates $\Gamma_1, \Gamma_2$ are calculated from the so-called PPT formulas \cite{Perelomov1966, Perelomov1967, Perelomov1967b}, while the single-photon ionization rates $\Gamma_3,\Gamma_4$ are obtained from experimental data \cite{Duncan2001}. Gain terms due to recombination processes (the positive analogs to the $\Gamma_j$ loss terms) are completely negligible on the sub-picosecond timescale that is studied here. 

Solving Eqs. \ref{schrodinger} to obtain the time evolution of the atomic state allows us to calculate the various terms on the RHS of Eq. \ref{waveeq}. The second term, which corresponds to atomic polarization due to transitions between bound states is:
\begin{equation}
\tilde{\mathcal{P}}= \mathfrak{F}\{\mathcal{N}\left(d_{21}a_1^*a_2 + d_{23}a_2^*a_3
+ d_{24}a_2^*a_4\right)\}.
\label{p}
\end{equation}
 This expression, together with Eqs. \ref{schrodinger} shows that: {\em i)} There is absorption in the medium due to single-photon transitions between bound states. These processes have considerable rates even at low 
intensity due to Rabi-oscillation type solutions of the equations. {\em ii)} Because the overall magnitudes of the probability amplitudes decrease due to the decay terms (loss of the valence electron during ionization), the induced atomic polarization decreases over time. Similar to atomic absorption that saturates when light is intense enough, the nonlinear polarization embodied in Eq. \ref{p} is thus {\em saturable}, it goes to zero as the valence electron detaches from the Rb$^{1+}$ core. {\em iii)} Besides the direct three-photon ionization from the ground state we have a two-photon ionization process from the first excited state and single-photon ionization from the two highest lying states. Because of the nonperturbative, Rabi-oscillation type solutions for the transitions between bound states, at low intensity the rates for these latter, combined processes 
(proportional to $\sim |a_2|^2 I^2$, $\sim |a_3|^2 I$ and $\sim |a_4|^2 I$) will surpass considerably the one for direct three-photon ionization $\sim |a_1|^2 I^3$. This means that the atoms are ionized much more easily by the resonant radiation. 

The third term on the RHS of Eq. \ref{waveeq} is purely an energy loss term
derived from the requirement that the laser pulse should lose an appropriate number of times the 
energy of a photon each time an atom is ionized:  
\begin{equation}
\tilde{\mathcal{Q}}=\mathfrak{F}\left\{\sum_j n_j 
\frac{\Gamma_j |a_j|^2}{\mathcal{E}^*}\right\}.
\label{q}
\end{equation}
$n_j$ are the number of photons taking part in the ionization process from state $|j\rangle$.
Finally the last term proportional to the ionization probability is the plasma dispersion term, with 
$\mathcal{R}$ proportional to the ionization probability:
\begin{equation}
\tilde{\mathcal{R}}=\mathfrak{F}\left\{
(1-\sum_j |a_j|^2)\mathcal{E}
\right\}
\label{r}
\end{equation}
The only difference between the present set of equations and those in \cite{Demeter2019} is the inclusion of the plasma dispersion, which in fact has very little effect as the vapor density is low and the complete conversion to Rb$^{1+}$ ions around the axis means that plasma density gradients appear only close to the edge of the pulse. One could also add a similar term due to plasma absorption but that would be orders of magnitude smaller as the electron collision rates are much less than the inverse pulse duration. The theory is valid for any inhomogeneous vapor distribution $\mathcal{N}(r,z)$, but we consider $\mathcal{N}$ constant here as experiments were performed with homogeneous vapor densities.

Note that the present theory contains optical nonlinearities due to resonant transitions between bound states, traditional nonresonant nonlinear optical coefficients are neglected. This approach can be justified by noting that  
medium polarization (linear or nonlinear) is proportional to the vapor density and in this case it is $10^4-10^5$ times smaller than the atmospheric density.
The critical power for self-focusing in air and atmospheric density gases is around or above the GW range \cite{Berge2007} for the ``standard'' nonresonant case, so they would be around or above the 10-100 TW range for our densities. The fact that the standard critical power formula is only sufficient for an order-of-magnitude estimate
for ultrashort pulses \cite{Polynkin2013} does not affect this estimate. Furthermore, there is no great difference between the nonlinear optical coefficients (hyperpolarizabilities) of $\mathrm{O_2}$, $\mathrm{N_2}$ and Ar, and only a factor of 2-3 difference between these and that of Kr \cite{Shelton1990}. So it is reasonable to expect that the nonresonant optical nonlinearities for rubidium can be neglected for $\sim$ 1 TW pulses in the present case. Note also that the theory is valid for ultrashort pulses, where the timescale is well below the ns timescale of atomic relaxation times. 

\section{Simulation results and comparison with experiment}

\subsection{Computer simulations}
\label{simulation}

In order to compare predictions of the theory with experimental data, a series of computer simulations were performed. 
The coupled equations \ref{waveeq} and \ref{schrodinger} with the relations \ref{p}, \ref{q} and \ref{r} were solved 
for an axisymmetric Gaussian input beam (TEM$_{00}$ mode, central wavelength $\lambda=780.241 \mathrm{~nm}$, duration $T=120 \mathrm{~fs}$, sech temporal pulse envelope for the electric field) that propagates in homogeneous rubidium vapor with density $\mathcal{N}$. 
Several sets of simulations were performed as detailed in table \ref{simpars}, sets A), B) and C) with parameters corresponding to the three sets of measurements with different vapor densities. Beam waist location $z_0$ (measured from the vapor entrance at $z=0\mathrm{~m}$) and waist radius parameter $w_0$ were determined by calculating the 
best fitting Gaussian beam to the three virtual laser line camera images for each single shot and using the average
of the best fit parameter values for each vapor density separately. Input energy was scanned in the experimental range and computed optical fields at the vapor exit $z=10\mathrm{~m}$ were used to determine the energy, spatial width and peak fluence of the transmitted pulse for comparison with the experiment. 

 \begin{table}[h]
 \begin{ruledtabular}
\begin{tabular}{|l|l|l|l|l|l|}
\hline
set label: & A) & B) & C) & D) & E) \\
\hline 
$\mathcal{N}~ [10^{14}\mathrm{~cm^{-3}}]$ & $1.87$ & 
$4.895$ & $6.6$ & $6.6$ & $6.6$\\

$z_0~[\mathrm{m}]$ & $6.11$ & $7.63$ & $7.92$ & $9.49$ & $6.35$\\

$w_0~[\mathrm{mm}]$ & $1.478$ & $1.507$ & $1.506$ & $1.517$ & $1.482$\\
\hline
\end{tabular}

\end{ruledtabular}
\caption{Simulation set parameters. Sets A)-C) correspond to best fits to experiment, sets D) and E) are perturbed parameter sets. Note that parameter perturbations in $w_0$ and $z_0$ are comparable in magnitude in the sense that if we have a Gaussian beam with $w_0 = 1.5\mathrm{~mm}$ and $z_R=9\mathrm{~m}$ (the value in our case), its waist  size $w(z)$ at $z=1.5\mathrm{~m}$ from focus is $w=1.52\mathrm{~mm}$. }
\label{simpars}
\end{table}

Additionally, simulations A)-C) were repeated in a series of ``null hypothesis'' calculations in an attempt to  
assess the importance of atomic resonances in the model. In these runs, the atomic model was reduced to contain only the ground state, resonant transitions to excited states were excluded. Eqs. \ref{schrodinger} were thus reduced to only the first one (for $a_1$, while $a_2=a_3=a_4=0$) and the second term on the right-handside of Eq. \ref{waveeq} is zero, only the diffraction, ionization loss and plasma terms remained.

\subsection{Comparison of simulation results with experimental data}

\begin{figure}[htb]
\includegraphics[width=0.35\textwidth]{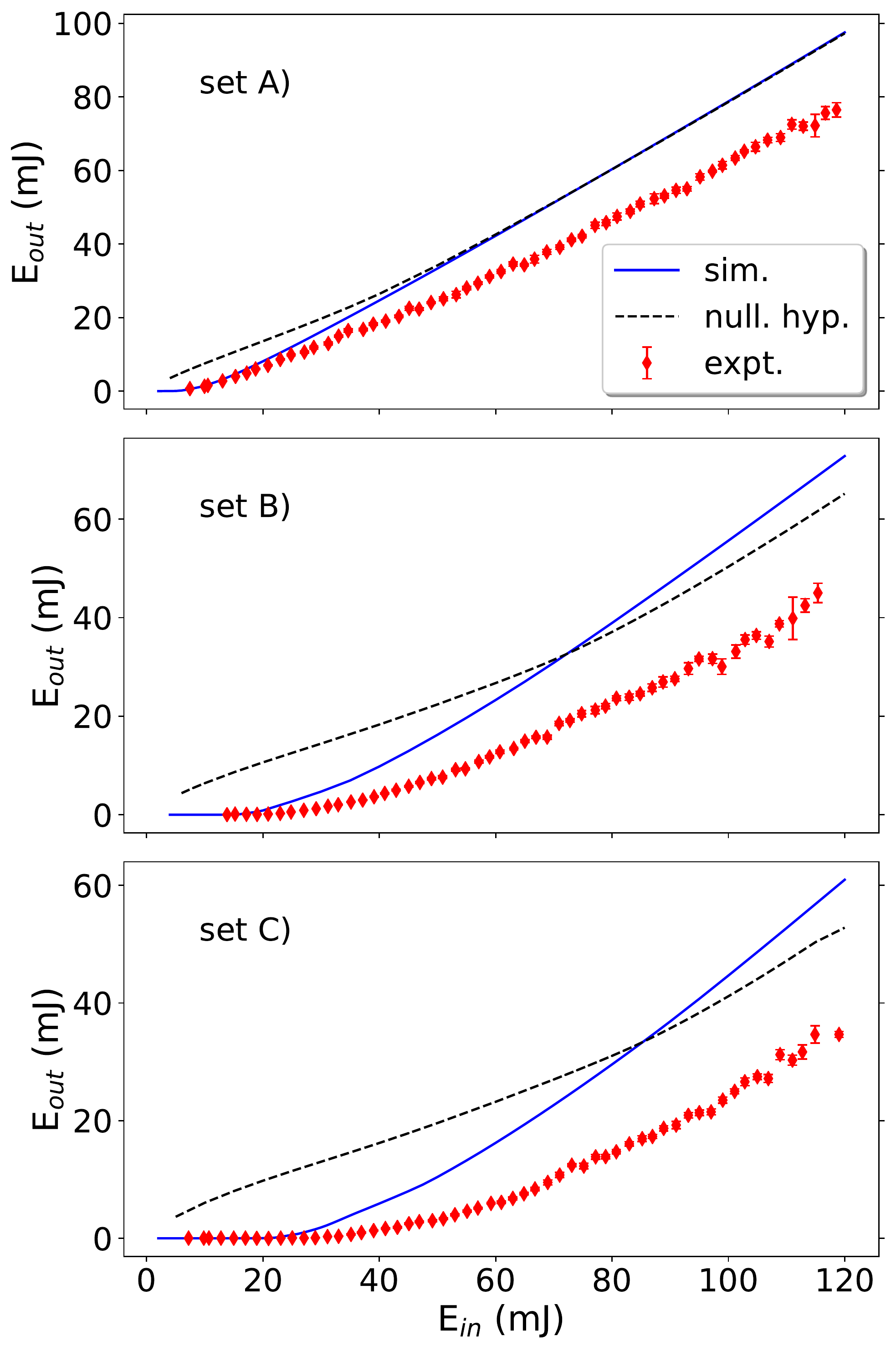}
\caption{Transmitted pulse energy vs. input pulse energy for the measurements and the corresponding simulations. Labels A)-C) correspond to the parameter set labels of table \ref{simpars}. Solid blue line: simulation with atomic resonances, dashed black line: reduced theory simulation (no resonances), red symbols: binned experimental data averages with error bars showing error of the mean (mostly smaller in size than the symbol marking the points).}
\label{energycurves}       
\end{figure}

Figure \ref{energycurves} shows the measured and calculated transmitted pulse energy $E_{out}$ as a function of input pulse energy $E_{in}$. Simulations clearly reproduce the breakthrough behavior observed ($E_{out}>0$ only above a certain threshold value of $E_{in}$), but predict lower threshold and higher transmitted energy above that 
(e.g. simulated breakthrough threshold for set C) is $\sim$24 mJ rather than the experimental $\sim$35 mJ, while maximum transmitted energy is 60 mJ instead of 35 mJ). The relative difference increases with vapor density. The reduced theory without resonances does not predict this breakthrough behavior, 
some energy is transmitted for arbitrary low input energies because in this case the medium is transparent when light intensity is too low for multiphoton ionization. (In fact, for very low laser pulse energy, for which multiphoton ionization is completely negligible, reduced theory would predict $E_{out} = E_{in}$, but for pulse energies plotted here, that is not the case because there is some ionization even at these low energies.) The agreement is therefore clearly better between experiment and the simulation results including resonances.

\begin{figure}[htb]
\includegraphics[width=0.35\textwidth]{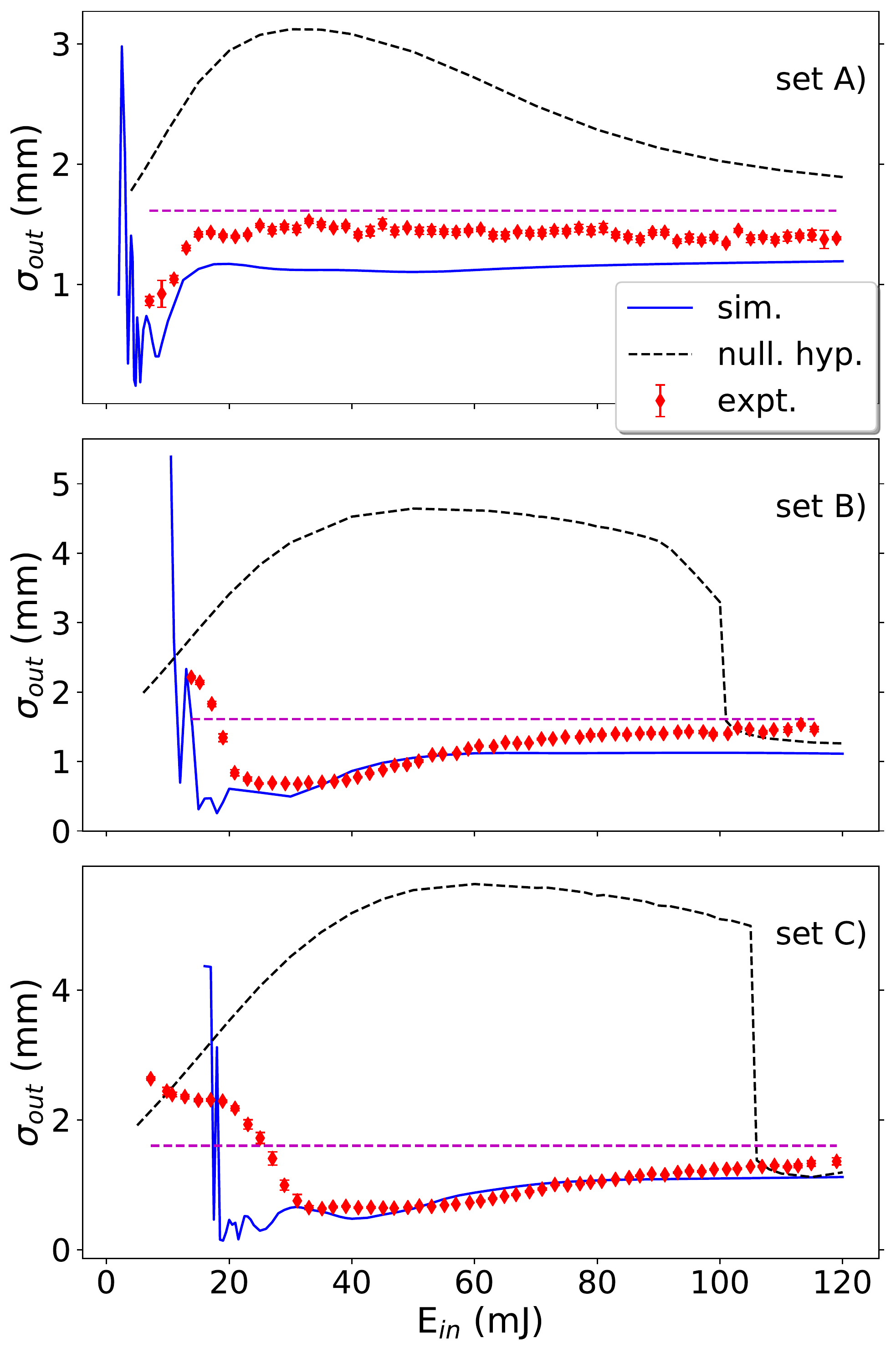}
\caption{Transmitted beam width vs. input pulse energy for the measurements and corresponding simulations.  Labels A)-C) correspond to the parameter set labels of table \ref{simpars}. Solid blue line: simulation with atomic resonances, dashed black line: reduced theory simulation (no resonances), red symbols: binned experimental data averages with error bars showing error of the mean (mostly smaller in size than the symbol marking the points). Dashed horizontal line marks the beam width measured on the virtual exit camera C3.}
\label{sigmacurves}       
\end{figure}

Figure \ref{sigmacurves} shows the Gaussian fit $\sigma$ of the transmitted beam fluence profile.  Whereas there is a fair qualitative similarity between the theoretical (with resonance) and experimental curves, reduced theory curves lie far from the former two. In particular, full theory exhibits a sharp drop in $\sigma$ around breakthrough and something similar to the confined beam domain just above it, but 
reduced theory does not. The steep drop in $\sigma$ occurs at smaller $E_{in}$ for simulation than for experiment, a feature also reflected in Fig. \ref{energycurves}.
We do note however, that the abrupt drops in output beam $\sigma$ for the calculated fluences may sometimes be  artificial, the real change in the shape of the energy distribution is not always so abrupt. 
The distribution can display shapes that are difficult to characterize with a Gaussian curve, e.g.  a superposition of a very narrow central peak on top of a wide background, or a distribution that is non-monotonic in $r$ (rings). In these cases, the fit parameters may exhibit abrupt jumps, e.g. when the fit starts favoring the central peak over the wide background at some point. The very sharp drops visible on the $\sigma$-curves of reduced theory on Fig. \ref{sigmacurves} B) and C) are such artifacts of the fit.  

\begin{figure}[htb]
\includegraphics[width=0.485\textwidth]{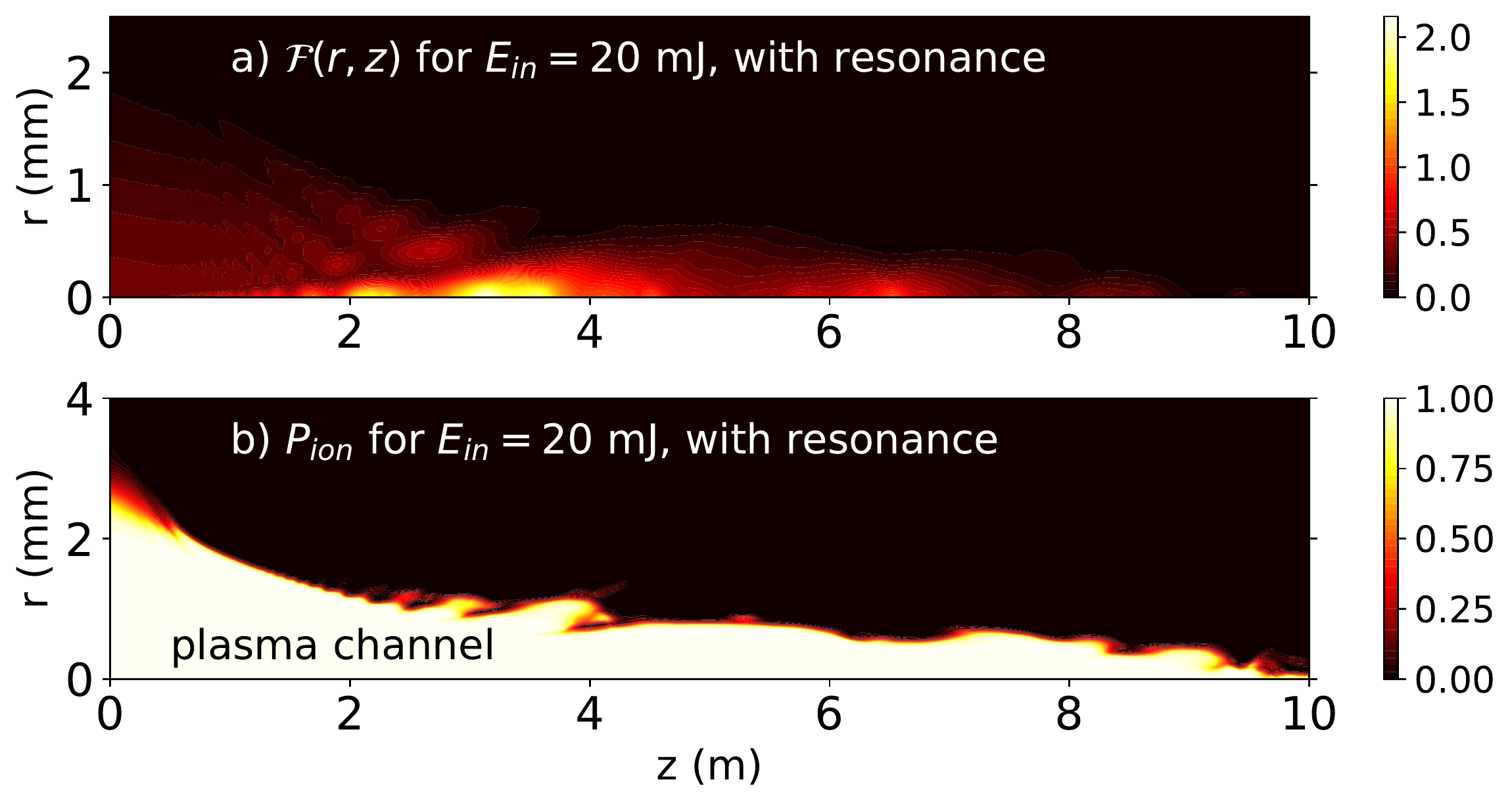}
\includegraphics[width=0.485\textwidth]{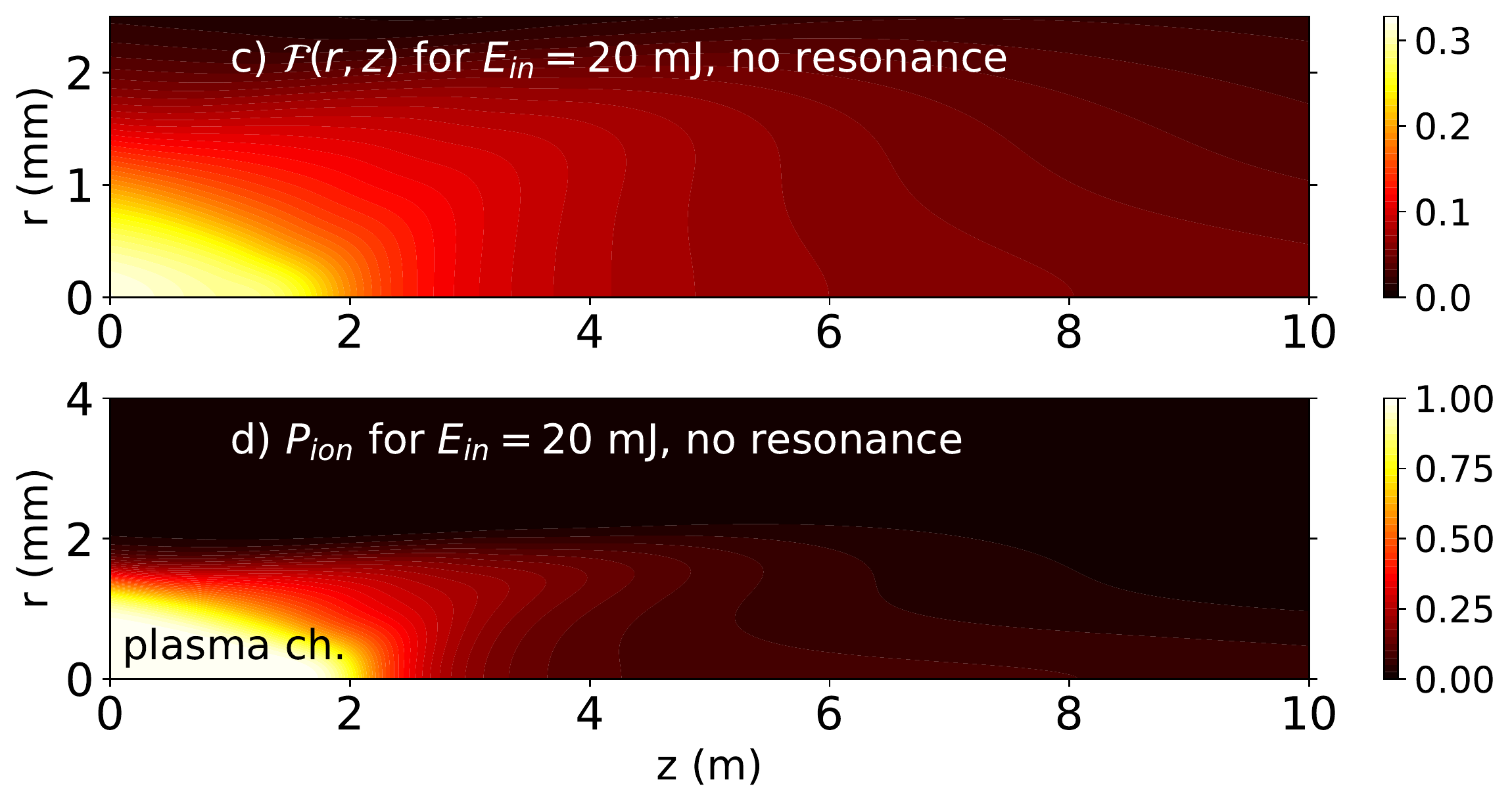}
\includegraphics[width=0.485\textwidth]{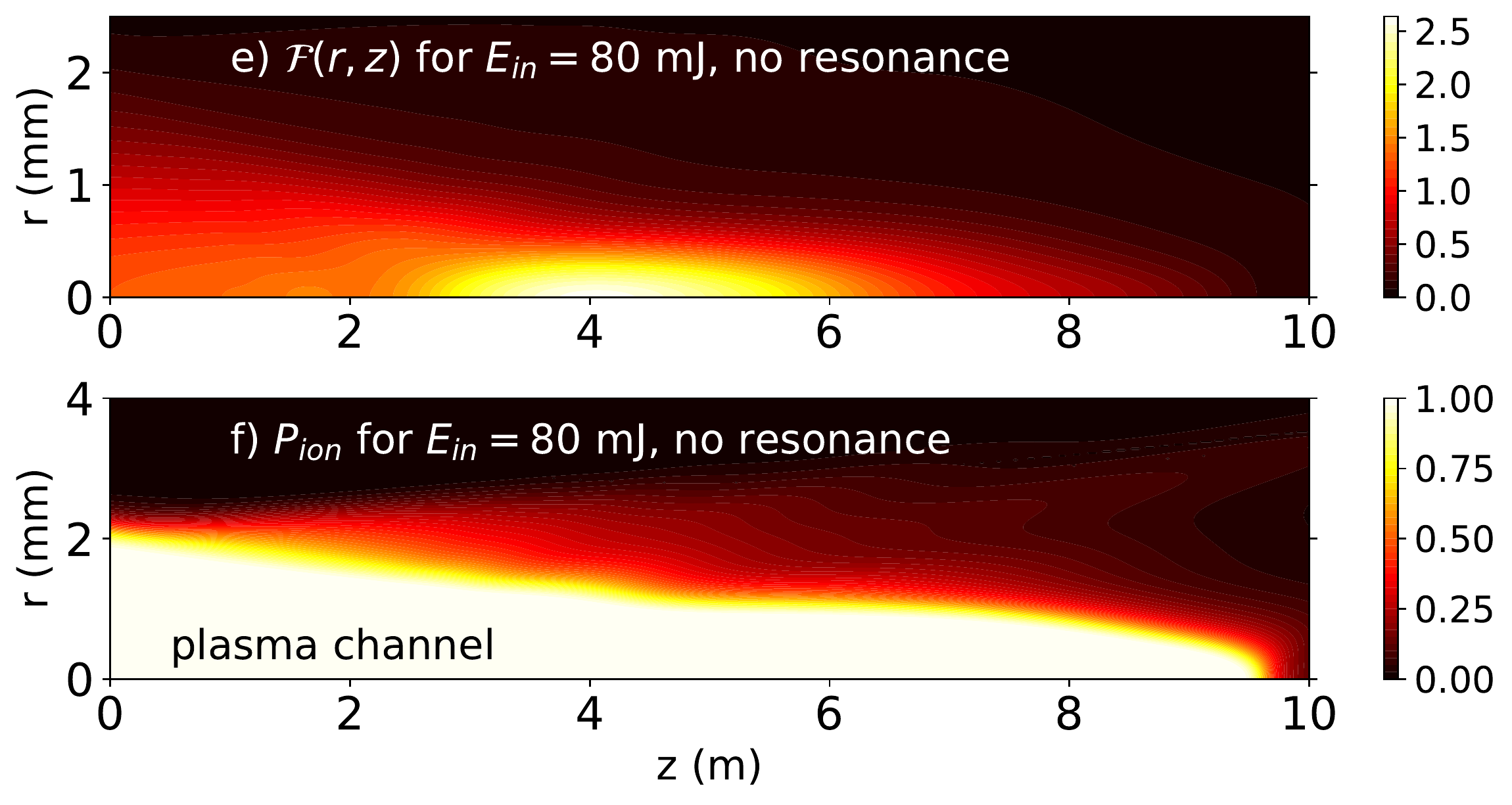}
\caption{Pulse fluence (in $\mathrm{J/cm^2}$) and atomic ionization probability profiles as a function of propagation distance $z$ and transverse radius $r$. a) $\mathcal{F}(r,z)$  and b) $P_{ion}$ from full theory with resonance $E_{in} =20 \mathrm{~mJ}$ pulse. c) $\mathcal{F}(r,z)$  and d) $P_{ion}$ reduced theory calculation, $E_{in} = 20 \mathrm{~mJ}$ pulse. e) $\mathcal{F}(r,z)$  and f) $P_{ion}$ reduced theory calculation, $E_{in} = 80 \mathrm{~mJ}$ pulse. Data was taken from simulation set C). 
`Plasma channel' marks area with complete conversion to Rb$^{1+}$ ions near axis. }
\label{propagcurves}       
\end{figure}

 {
To illuminate the difference between predictions of the full theory and reduced theory, we plot the calculated fluence and ionization profiles (i.e. the extent of the plasma channel) in space for various  pulses in both cases on Fig. \ref{propagcurves} a)-f). 
One important difference visible is the long, narrow beam with repeated self-focusing maxima of a $E_{in}=20\mathrm{~mJ}$ pulse predicted by full theory (Fig. \ref{propagcurves} a) ), whereas the beam is 
much wider for the same pulse when calculated using reduced theory (Fig. \ref{propagcurves} c) ).
The corresponding plasma channel with complete conversion to Rb$^{1+}$ ions that was calculated using full theory is much longer, almost reaching the downstream end of the vapor, with an oscillating radius and very sharp boundary (Fig. \ref{propagcurves} b) ), whereas it is short for reduced theory with
a wide transition region of partially ionized vapor (Fig. \ref{propagcurves} d) ). According to reduced theory, it takes a pulse of much higher energy, $E_{in}=80\mathrm{~mJ}$ to produce a plasma channel with complete conversion to Rb$^{1+}$ ions that is about as long as the one with $E_{in}=20\mathrm{~mJ}$ in the resonant case (Fig. \ref{propagcurves} f) ). 
The reduced theory calculation exhibits a single fluence maximum due only to the Gaussian beam waist  (Fig. \ref{propagcurves} e) ) for this large energy pulse. Transmitted energy is $E_{out}\approx 0$ for full theory calculation, whereas it is $E_{out}\approx 9.8 \mathrm{~mJ}$ and  $E_{out}\approx 31 \mathrm{~mJ}$ for the reduced theory for the two initial pulse energies shown. }
  
\begin{figure}[htb]
\includegraphics[width=0.35\textwidth]{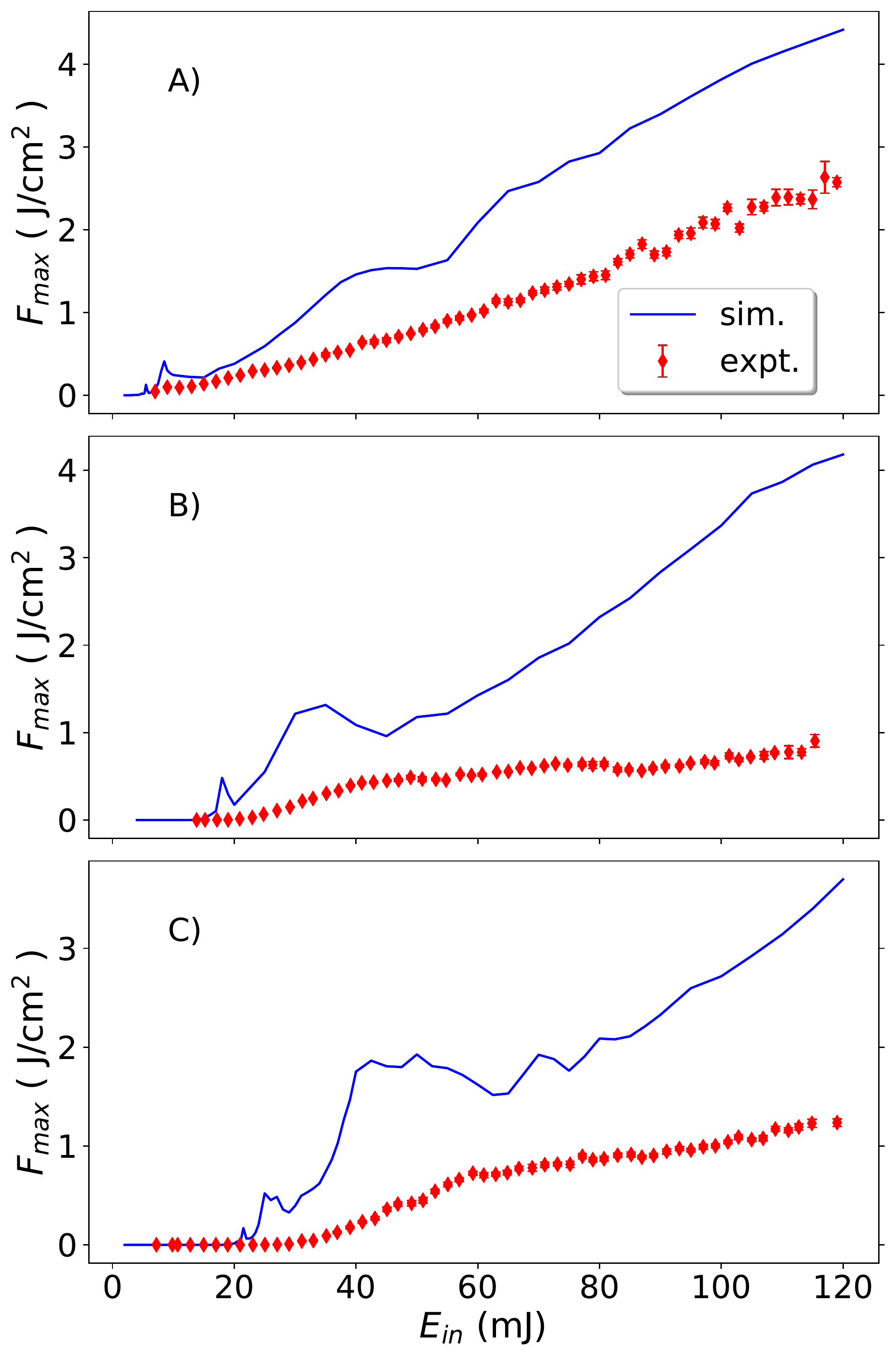}
\caption{Transmitted pulse on-axis fluence in $\mathrm{J/cm^2}$ as a function of input energy for the three simulation series. Labels A)-C) correspond to the parameter set labels of table \ref{simpars}. Solid blue line: simulation, red symbols: binned experimental data averages with error bars showing error of the mean (mostly smaller in size than the symbol marking the points).}
\label{onaxis_fluences}       
\end{figure}

Finally, Fig. \ref{onaxis_fluences} shows the predicted on-axis fluence values with the experimental data. For the two larger densities ( B) and C) ), where the experimental data shows a steep increase of on-axis fluence initially, followed by slower increase (corresponding to growth during and above the confined beam region), the simulated curves show a much steeper increase. The relative difference is much larger than the difference between the transmitted energy (Fig. \ref{energycurves}). The two regions of different slopes can nevertheless be recognized for the highest density calculation Fig. \ref{onaxis_fluences} set C).

\subsection{Pulse parameter variability}

One feature visible on Fig. \ref{sigmacurves} is the fact that where the experiment captures the confined beam region just above breakthrough, simulation does not predict a constant exit beam $\sigma$, but a series of oscillations before a monotonous increase.  
The oscillatory nature of $\sigma$ with the pulse energy just above breakthrough in the simulation is easily understood by looking at Fig. \ref{propagcurves} a)-b), which show that during propagation, the laser pulse experiences repeated self-focusing phases with  oscillatory on-axis fluence, transverse width and plasma channel radius values along the propagation axis $z$. Laser pulses with different parameters (in particular, different $E_{in}$) exhibit oscillations that are identical in nature, but locations along the $z$ axis of fluence or beam width maxima or minima vary considerably. This translates into oscillations in values observed at $z=10\mathrm{~m}$ as the laser pulse energy is varied.

\begin{figure}[htb]
\includegraphics[width=0.485\textwidth]{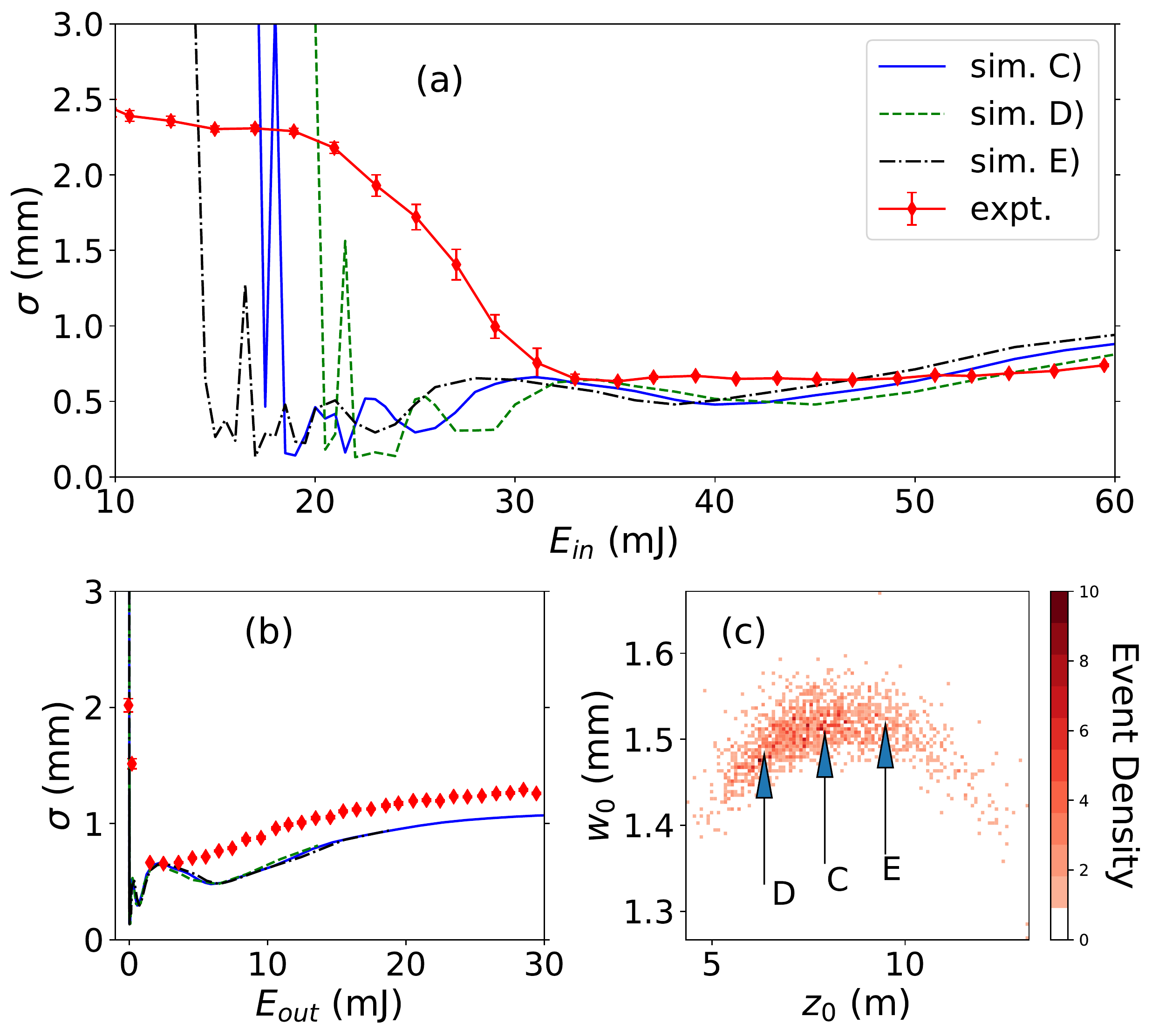}
\caption{(a) Transmitted beam width vs. input pulse energy for the original simulation C) and the two simulations with modified Gaussian beam parameters D) and E). Binned experimental data with error of the mean is also plotted. (b) The same quantities plotted with respect to the transmitted pulse energy. (c) 2D histogram of the input beam parameters obtained for $\mathcal{N}=6.6\cdot 10^{14}\mathrm{~cm}^{-3}$ density shots with arrows pointing to the mean values C) and perturbed parameter simulation values D) and E).  }
\label{width_values}       
\end{figure}

The quantitative comparison of simulation and experiment is hampered by the fact that the axisymmetric Gaussian beam and constant beam parameters $z_0$, $w_0$ used in the calculations do not model the experimental situation very well. First, the laser beam exhibits considerable ellipticity. To quantify this, an elliptically symmetric Gaussian function was used in a second fit on the virtual exit camera (C3) images that contained two width parameters ($\sigma_{max}$ and $\sigma_{min}$) and an $\alpha$ angle parameter that determined the orientation of the ellipse major axis in the $x-y$ plane. Calculating the ellipticity parameters for the fits we obtain a mean value of $f = 0.297\pm0.015$. Second, the Gaussian beam fit to the virtual laser line images that is used to obtain the input beam parameters for the simulations exhibits considerable shot to shot fluctuations of the parameters. To check the corresponding variability of the simulation results, we performed two additional series of simulations, with perturbed beam parameters (series D) and E) in table \ref{simpars}). The parameters were selected to be representative of the variation of the set of beam parameters - a 2D histogram of the set of input beam parameters and the selection of simulation parameters are shown in Fig. \ref{width_values} (c).

The transmitted pulse $\sigma$ obtained using the perturbed parameter simulations can be seen on Fig. \ref{width_values} (a), together with experimental data and the original simulation set C). One can see that the precise location of the sudden drop in transmitted beam width associated with the breakthrough, as well as the location of the width minima and maxima just above it show considerable variation with the beam parameters. 
In fact, the variation in the location of the large drop in beam width from simulation is about the same size as the extent of the breakthrough domain with large beam width fluctuations on the experimental data plots. This strongly suggests that it is primarily the input beam parameter fluctuations that define the extent of this  domain along the $E_{in}$ axis. The oscillatory beam width predicted by simulation above breakthrough is expected to be 'washed out' due to beam parameter fluctuations in the experiment, as the typical variation in beam parameters yields maxima and minima at different places along the propagation axis.

Figure \ref{width_values} (b) shows the same curves, this time plotted with respect to transmitted pulse energy $E_{out}$. The plot shows, that simulated $\sigma$ curves are now in phase with respect to each other, i.e. transmitted beam properties correlate much more directly with $E_{out}$. They also follow much better the experimental trend for $E_{out}\gtrsim 5\mathrm{~mJ}$, than on Fig. \ref{width_values} (a), though there is still a constant shift (simulated beams are narrower) and a local maximum for very small $E_{out}$. It is likely that these differences can be attributed to experimental beam ellipticity and higher order spatial mode content, not taken into account in the simulation. The sub-threshold and breakthrough domains are naturally squeezed around the origin on Fig. \ref{width_values} (b) and not visible.

\section{Discussion}

As demonstrated, the theory that includes an explicit treatment of the resonant atomic bound states for the calculation of the nonlinear optical response shows qualitative agreement with experimental observations, whereas the null-hypothesis theory where this is missing, does not. This proves that it is essentially correct to include the transient atomic response in the propagation equation and that single-photon resonances do indeed play the dominant role in this setting. We also conclude that the calculations can be used to interpret the qualitative behavior observed and obtain information on the properties of pulse propagation inside the vapor cell where we can make no measurements. Here, we briefly summarize some key features of the pulse propagation that can be inferred from the simulation results. A more complete account can be found in \cite{Demeter2019}. 

\subsection{Pulse evolution during propagation}

The self-focusing of the beam is evident on Figs. \ref{propagcurves} a) and b) - at the same time
Eqs. \ref{waveeq} and \ref{schrodinger}, encountered in {\em resonant nonlinear optics} are substantially different from standard equations in nonlinear optics where the material response is derived from susceptibility functions of increasing order. Self-focusing in this system takes place via {\em coherent on-resonance self-focusing} \cite{Gibbs1976, Lamare1994}, which is a fundamentally different process from traditional self-focusing caused by an intensity dependent refractive index. The plane wave (1D) on-resonance propagation problem in a two-level medium gives rise to the classical secant-hyperbolic Self-Induced Transparency (SIT) solutions \cite{McCall1969}. Here the important quantity is the pulse area, which is proportional to the time integral of the pulse amplitude. Pulses entering the medium are either absorbed or reshape to $2\pi$ area pulses (or a sequence of distinct $2\pi$ pulses if the initial area is high enough). These pulses then propagate without further attenuation or distortion in the plane wave limit with a speed depending on the pulse duration (slow light). For a field that varies in a radial direction, each annular region produces a SIT soliton (or sequence of solitons) with different duration and hence different velocity. Overall, this leads to the distortion of the phase front of the original pulse and eventually self-focusing. The properties of this type of self focusing (e.g. the threshold of the onset, the focusing distance and its dependence on the initial pulse diameter) differ from those of the traditional self-focusing process  \cite{Lamare1994}.

\begin{figure}[htb]
\includegraphics[width=0.485\textwidth]{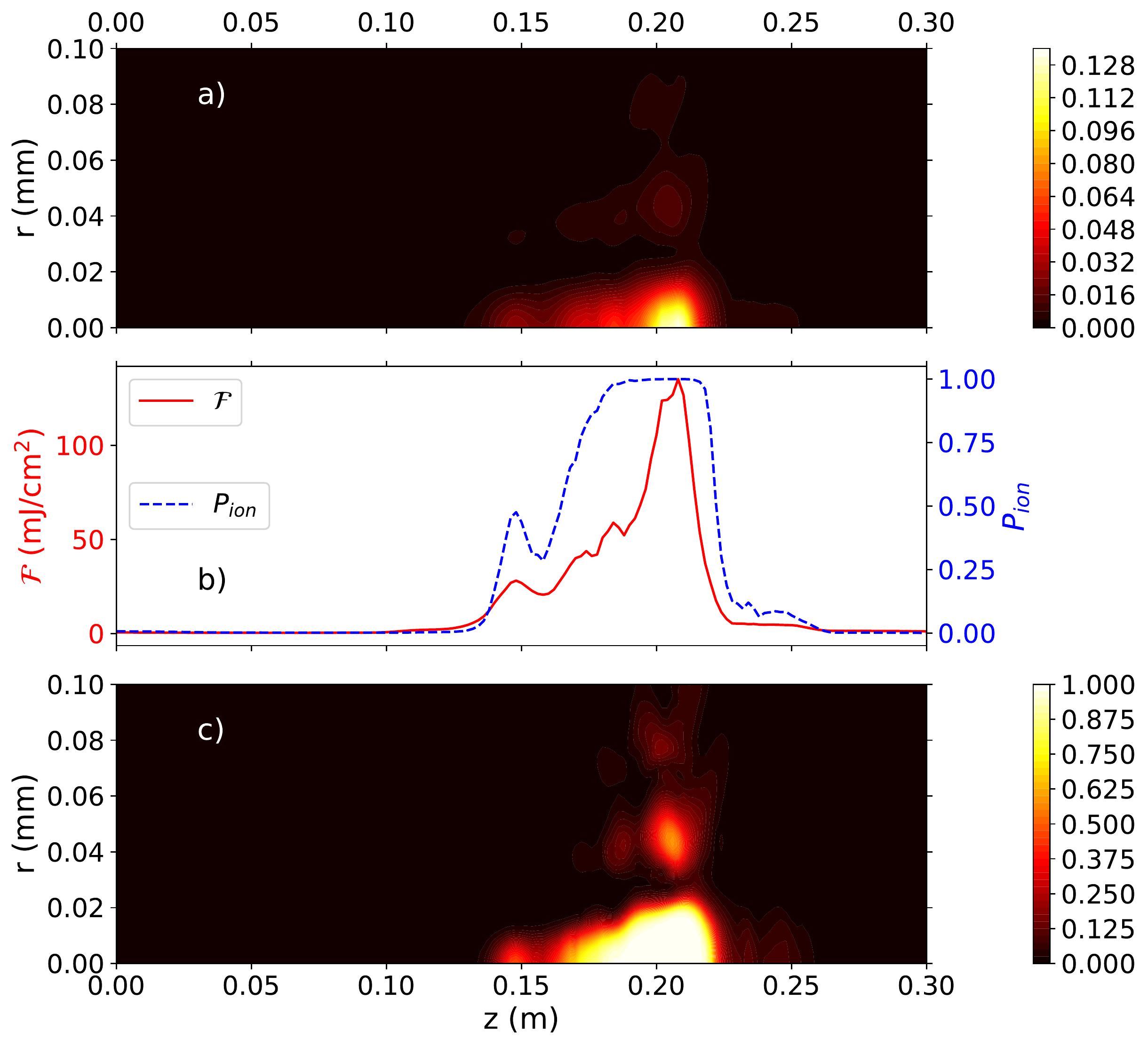}
\caption{The propagation and self-focusing of a low energy, $E_{in}=0.04\mathrm{~mJ}$ pulse of parameter set C). a) fluence $\mathcal{F}(r,z)$, b) on-axis fluence $\mathcal{F}(r=0,z)$ and ionization probability $P_{ion}(r=0,z)$ and c) ionization probability $P_{ion}(r,z)$.}
\label{self_focusing}       
\end{figure}

In our system ionization when the pulse intensity becomes high enough and the two higher lying excited states cause further complications. However, for a low energy pulse, the intensity is initially small enough for the system to behave as a two-level medium. As peak intensity grows during propagation due to self-focusing, transitions to higher lying excited states and ionization start and the pulse deposits its remaining energy in a relatively short distance. Figure \ref{self_focusing} depicts the evolution of a $E_{in}=0.04\mathrm{~mJ}$ pulse in simulation set C). The 
fluence and the ionization profiles show that the pulse self-focuses and at around $z=0.2$ m has a diameter of around 40 $\mu$m. Ionization is restricted to the immediate vicinity of the focus. By contrast, the $E_{in}=20\mathrm{~mJ}$ pulse depicted in Fig. \ref{propagcurves} a) and b) is intense enough to ionize from the very start, experiences a series of focusings in the medium and it is not focused to such a narrow beam diameter, except at the very end where pulse energy has been almost completely depleted. 

\begin{figure}[htb]
\includegraphics[width=0.485\textwidth]{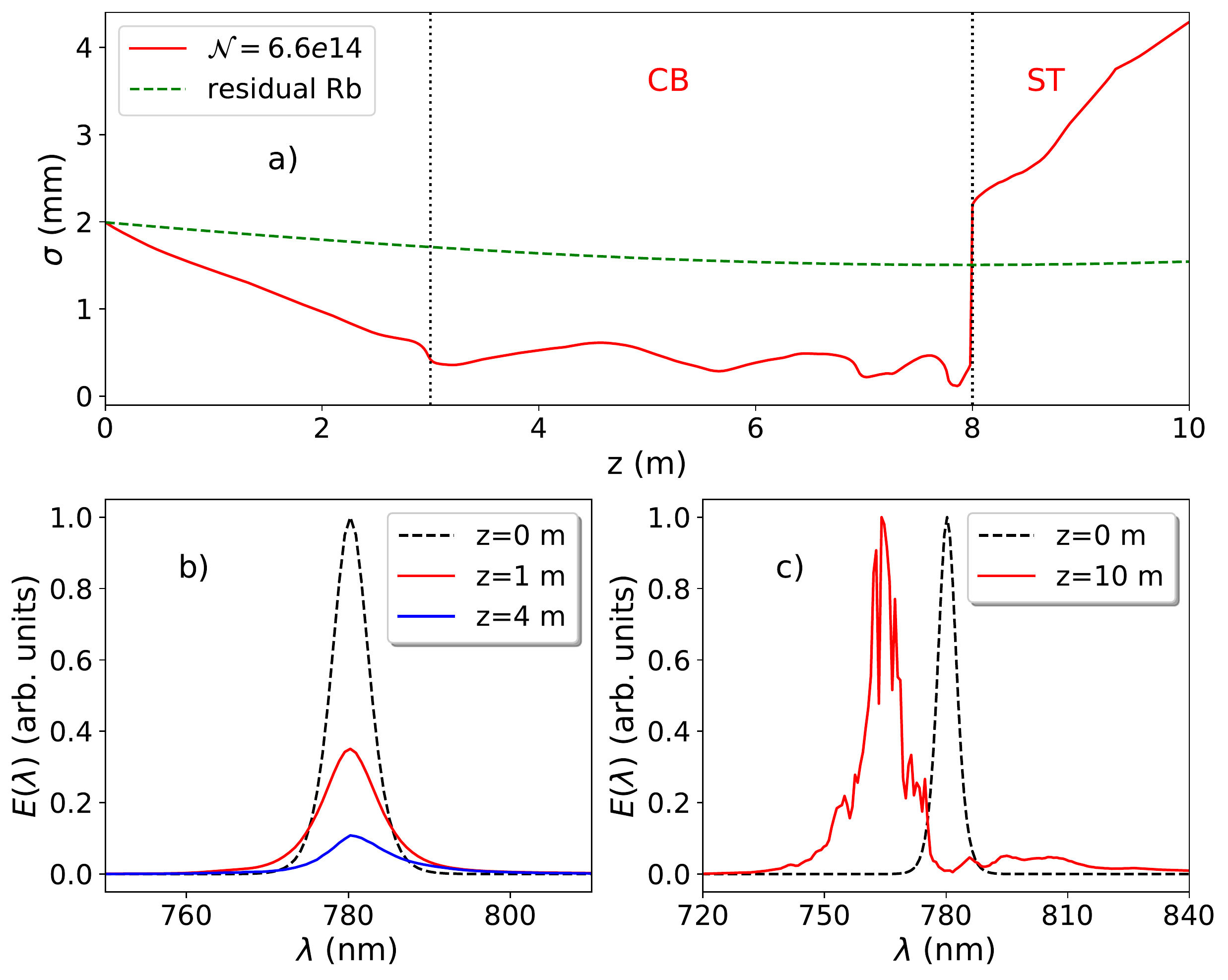}
\caption{The propagation of a $E_{in}=16\mathrm{~mJ}$ pulse for simulation parameter set C). a) Beam width $\sigma$ as a function of propagation distance $z$. Vertical dotted lines delimit the regions that can be associated with the confined-beam (CB) and sub-threshold (ST) domains. Dashed line marks the width parameter $w(z)$ for the unperturbed Gaussian beam.  b) Pulse energy spectrum at the vapor entrance $z=0$ m and at $z=1$ m inside the vapor, drawn to scale. c) Normalized pulse energy spectrum at the two ends of the vapor source, $z=0$ m and $z=10$ m.  }
\label{propag_spectrum}       
\end{figure}

Figure \ref{propag_spectrum} depicts plots of a $E_{in}=16\mathrm{~mJ}$ pulse in simulation set C). This pulse is intense enough to ionize atoms already at the start of the vapor source, but is not energetic enough to do so all the way to the downstream end. The evolution of the beam width $\sigma$ along the propagation direction $z$ is shown in Fig. \ref{propag_spectrum} a). The beam first contracts in an initial focusing regime (until $z\approx3$ m) after which the the beam width starts to oscillate with repeated self-focusing phases. The average beam width changes relatively little in this regime, so we can readily associate this region of propagation with the confined-beam domain. At $z\approx 8$ m the beam width abruptly increases and becomes much wider than that of the same Gaussian beam propagating in residual vapor (shown by the dashed line). This transition can clearly be associated with the breakthrough transition discussed earlier, i.e. a 16 mJ pulse would be just around breakthrough at the end of an 8 meter vapor source with these density and beam parameters. Above $z\approx 8$ m, the propagation can be associated with the sub-threshold domain. Calculations show that the plasma channel with full conversion to Rb$^{1+}$ ions stretches almost to the point where the sudden increase in width is observed, so the term ``breakthrough'' can be interpreted as the approximate point where the plasma channel with full conversion to Rb$^{1+}$ ions reaches the downstream end of the vapor.

Figure \ref{propag_spectrum} b) and c) depict the changes in pulse energy spectrum relative to the initial spectrum. The one after a propagation distance of $z=1$ m (drawn to scale with the initial spectrum) shows a widening of the spectrum on both the blue and the red side. The final spectrum at the downstream end of the vapor (normalized spectra presented as the output energy is a very small fraction of the input energy) shows the central, 780 nm components fully absorbed and a considerable blue-shifted peak present.

\subsection{Possible causes of quantitative discrepancy}

The substantial quantitative discrepancies between theory and experiment, especially for transmitted pulse energy and peak fluence prove that the current version of the model has limited predictive power, some points still need considerable refinement. It is probable that the discrepancies cannot be attributed solely to the difference between ideal simulated Gaussian beam and real experimental beam properties.

One additional cause is probably the overly simplistic description of ionization employed in the theory.
The description with intensity dependent ionization rates could be inaccurate as
the PPT formulas, derived with the assumption that the multiquantumness parameter is large $K_0 = E_{ionization}/\hbar\omega_0 \gg 1$ may have a limited validity for the three- and two-photon ionization processes of our case, especially for high intensities. A recent investigation of rubidium ionization \cite{Wessels2018} demonstrated  that ab initio calculations were needed to achieve quantitative agreement with experiment, especially when light is resonant with transitions between bound states. The wavelengths studied are different from the 780 nm in this investigation, pulse durations are much longer and single-photon resonances were not studied. In \cite{pocsai2019} an ab initio calculation of the ionization of rubidium atoms is presented that shows the appearance of above threshold ionization peaks in the emitted electron spectrum for peak pulse intensities already around $10^{12}\mathrm{~W/cm^2}$. In our case the peak intensity of a 100 mJ pulse at the focus would exceed $10^{13}\mathrm{~W/cm^2}$ with no vapor in the source. This could possibly explain the enhanced energy loss observed in the experiment when compared to our theory. 
However, the calculation in \cite{pocsai2019} has been done for a slightly different wavelength (800 nm). Furthermore, it predicts that there is a plateau for the ionization probability around 0.95, implying that there is a small fraction of the atoms that are not ionized even if peak intensities reach $10^{14}-10^{15}\mathrm{~W/cm^2}$. This prediction
does not seem to agree with observations at AWAKE, where the plasma density inferred from proton beam modulation suggests that plasma density equals the vapor density with an accuracy of 1\% \cite{Adli2019}. 

Inaccuracy may also be caused by using an ideal sech pulse time envelope in the simulation. Comparison of the spectrum of the ideal simulated pulse (Fig. \ref{propag_spectrum} b), dashed black line) and the measured spectrum of the laser (Fig. \ref{fig_Rblevels} b) ) reveals that the latter is much broader and different in shape.

Another possible source of discrepancies may be the reduction of the theory to a four-level system. While we included states with transitions within the initial spectrum of the laser pulses, high field amplitudes may Stark-shift other, previously nonresonant states into resonance as well.

\subsection{Further comments}

According to the observations presented, the confined beam region is the one that is the most similar to traditional laser beam filamentation, with the emerging beam width being constant over an interval of the pulse energies. 
However, this energy range is fairly narrow, the vapor cannot maintain the constant beam width for high-energy pulses because the nonlinearity responsible is saturable, the medium becomes transparent when all atoms are converted to Rb$^{1+}$ ions. According to theory \cite{Demeter2019}, the laser pulse energy propagates in the central plasma channel. This is unlike traditional filamentation where most of the energy propagates in the low intensity wings of the pulse, with absorption becoming significant in the high-intensity center. 
Contrary to this, in our case the high-intensity part of the pulse quickly renders the vapor transparent while there is always absorption in the low-intensity wings. 

Our investigations were focused on the specific case of rubidium, but it is probable that similar scenarios could be observed for other alkali atoms as well, where the outermost electron has an ionization potential much less than the energy needed to remove the second electron. Ionization potentials are quite similar for Li (5.39 eV), Na (5.14 eV), K (4.34 eV) and Cs (3.89 eV) and all atoms possess a strong optical resonance between the ground state and the first excited state such that ionization of the first electron requires three photons. 

Finally we note that the separation of the beam into multiple filaments as seen for high power laser pulses in dense atmospheric gases seems to be largely absent in the present case. Clear, multiple peaked distributions were observed only in a few cases, for relatively small energies around breakthrough (see Fig. \ref{pickoff_width}, inset c) ) and not for pulses with higher energy. The probable cause is that in the resonant setting, beam breakup occurs when the central area of the beam has a large pulse area, several times $2\pi$. When ionization is taken into account, the {\em effective} area of the pulse is reduced because the strong resonant interaction between field and atoms ceases.  

\section{Summary and outlook}

To summarize, we have studied the long range propagation of an ultrashort, ionizing laser pulse in rubidium vapor under conditions of single photon resonance from the atomic ground state and also between excited state transitions. Experiments were performed at the CERN AWAKE site and results compared to computer simulations of the propagation. Experiment and theory agree qualitatively and suggest that the model is useful in interpreting the observed phenomena. Pulse breakthrough was observed when the laser pulse was energetic enough to achieve single electron ionization of all atoms along the propagation axis, and a confined beam domain was identified just above that, where the width of the emerging laser pulse was approximately constant. 

Because of the quantitative differences between theory and experiment, we are planning further experiments to better determine the main cause(s) of the discrepancy and to better understand the interaction between the vapor and the laser pulse. Propagation experiments are foreseen with simultaneous measurement of the transmitted laser pulse spectrum, as well as possible Schlieren imaging of the plasma channel in a transverse direction near the end of the vapor source. A set of experiments and simulations with the spectrum of the ionizing laser pulse shifted away from resonance is also planned to better explore the importance of resonant interaction. Measurements of rubidium ionization with wavelengths close to resonance with a transition from the ground state are also planned, as the accuracy of the model could possibly be improved significantly by including more atomic levels in the model and a better description of the ionization process. The nature of the transverse modulations the beam may experience around breakthrough will also be studied further.

These results are important for the AWAKE experiment that aims at driving wakefields in the plasma for particle acceleration \cite{Adli2018}. For this application the plasma column radius must exceed the plasma skin depth (e.g., $207\mathrm{~\mu m}$ at a density of $6.6\cdot 10^{14}\mathrm{~cm}^{-3}$) over the entire plasma length. Results of the simulation suggest that this is realized already above breakthrough, in the confined beam domain, i.e. for $E_{in}\gtrapprox 36\mathrm{~mJ}$. However, the `safe' regime of operation that a particle acceleration project can rely on is clearly the asymptotic transparency domain. 
Once sufficient quantitative agreement is achieved between theory and observations, the calculation method presented here will be used to determine for example over what distance a large enough plasma radius can be formed as a function of laser pulse energy and vapor density.

\begin{acknowledgments}
The support of the National Office for Research, Development and Innovation (NKFIH) under contract numbers 2019-2.1.6-NEMZ\_KI-2019-00004 and 2018-1.2.1-NKP-2018-00012 is gratefully acknowledged.
The use of the Wigner Datacenter Cloud facility was indispensible for the numerical 
computations and its use through the {\em Awakelaser} project is gratefully acknowledged. We thank P. 
L\'{e}vai for his support.

\end{acknowledgments}

\clearpage

\bibliography{/home/gdemeter/fiz/manuscript/bibliography_pulseprop}

\bibliographystyle{apsrev4-1}


\end{document}



\title{Supplemental material: Long range propagation of ultrafast, ionizing laser pulses in a resonant nonlinear medium}%

\author{G. Demeter}
\email{demeter.gabor@wigner.hu}
\affiliation{Wigner Research Centre for Physics, Budapest, Hungary}

\author{J. T. Moody}
\affiliation{Max Planck Institute for Physics, Munich, Germany}

\author{M. \'{A}. Kedves}
\affiliation{Wigner Research Centre for Physics, Budapest, Hungary}

\author{B. R\'{a}czkevi}
\affiliation{Wigner Research Centre for Physics, Budapest, Hungary}

\author{M. Aladi}
\affiliation{Wigner Research Centre for Physics, Budapest, Hungary}

\author{A.-M. Bachmann}
\affiliation{Max Planck Institute for Physics, Munich, Germany}

\author{F. Batsch}
\affiliation{Max Planck Institute for Physics, Munich, Germany}

\author{F. Braunm\"{u}ller}
\affiliation{Max Planck Institute for Physics, Munich, Germany}

\author{G. P. Djotyan}
\affiliation{Wigner Research Centre for Physics, Budapest, Hungary}

\author{V. Fedosseev}
\affiliation{CERN, Geneva, Switzerland}

\author{F. Friebel}
\affiliation{CERN, Geneva, Switzerland}

\author{S. Gessner}
\affiliation{CERN, Geneva, Switzerland}
\affiliation{SLAC National Accelerator Laboratory, Menlo Park, California, USA}

\author{E. Granados}
\affiliation{CERN, Geneva, Switzerland}

\author{E. Guran}

\author{M. H\"{u}ther}
\affiliation{Max Planck Institute for Physics, Munich, Germany}

\author{V. Lee}
\affiliation{University of Colorado Boulder, Colorado, USA}

\author{M. Martyanov}
\affiliation{Max Planck Institute for Physics, Munich, Germany}
\affiliation{CERN, Geneva, Switzerland}

\author{P. Muggli}
\affiliation{Max Planck Institute for Physics, Munich, Germany}

\author{E. \"{O}z}
\affiliation{Max Planck Institute for Physics, Munich, Germany}

\author{H. Panuganti}
\affiliation{CERN, Geneva, Switzerland}

\author{L. Verra}
\affiliation{Max Planck Institute for Physics, Munich, Germany}
\affiliation{CERN, Geneva, Switzerland}
\affiliation{Technical University Munich, Munich, Germany}

\author{G. Zevi Della Porta}
\affiliation{CERN, Geneva, Switzerland}
 

\date{\today}

\begin{abstract}
We present below some additional information on the measurement setup and the calibration procedure of the energy meters.
\end{abstract}

\pacs{}
\maketitle

\section{Experimental arrangement}

\begin{figure*}[p]
\includegraphics[width=0.98\textwidth]{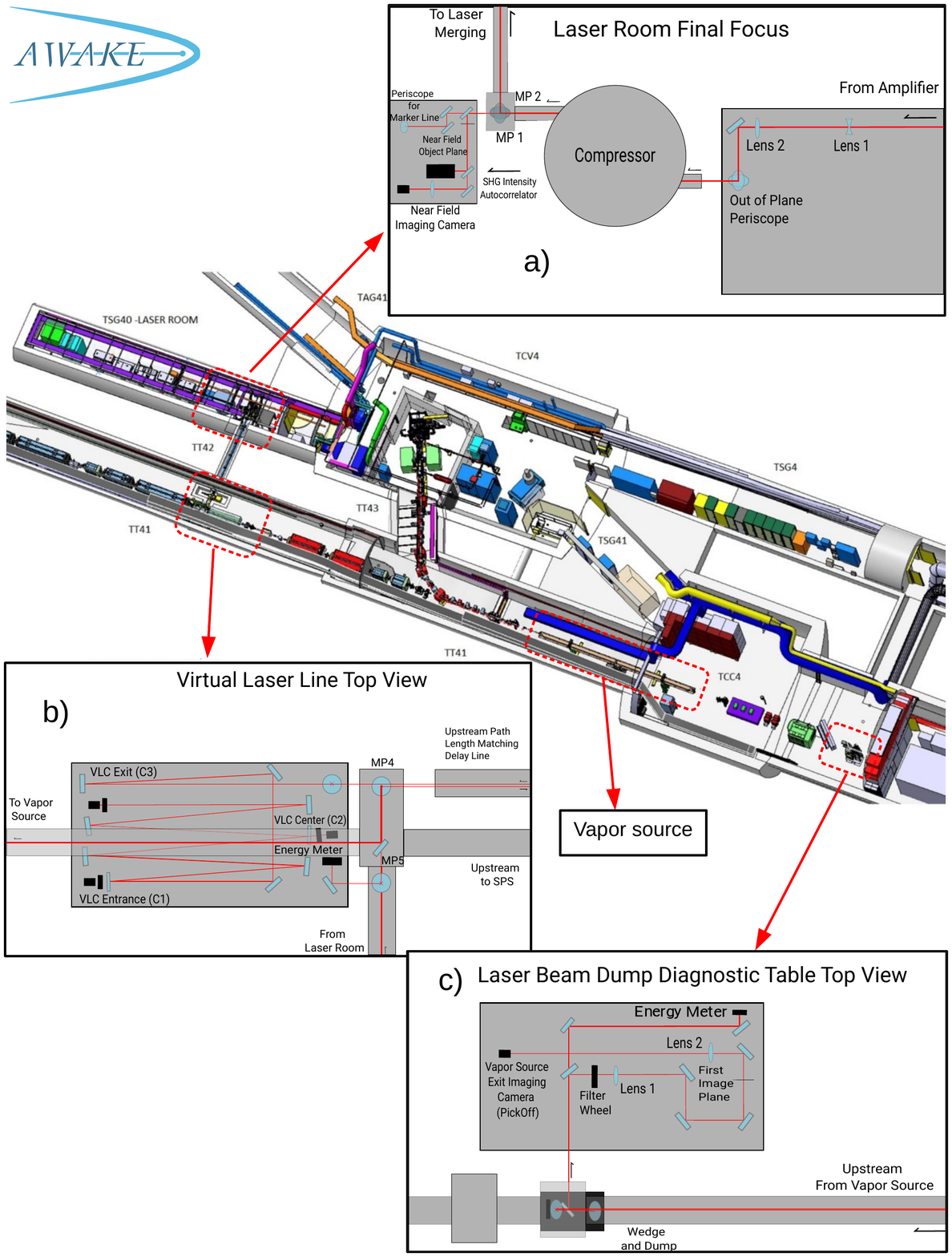}
\caption{ Figure of the AWAKE experimental area at CERN with the location of the various components of the experiment indicated and details of each area shown in the subfigures. a) Laser room with compressor and final focus. b) Laser and proton beam merging point with the virtual laser line setup. c) Laser beam dump diagnostic table with the pickoff camera and output energy meter.   }
\label{exp_setup}       
\end{figure*}

Figure \ref{exp_setup} depicts the AWAKE experimental area at CERN. Components of our experimental setup placed at various locations are encircled on the figure and subfigures indicate a detailed setup of each component. Each of the subfigures is explained in detail in the text below.

\subsection{Laser room final focus}

The 300 ps FWHM duration ionizing laser pulse leaves the main amplifier with a maximum of 600 mJ with a FWHM diameter of 13 mm. For the purposes of avoiding damage in the compressor, this energy is currently limited to 200 mJ.

The ionizing laser pulse next enters a beam expanding telescope. The telescope not only expands the FWHM to approximately 30 mm, but the lenses are slightly separated beyond the confocal such that the pulse has approximately milliradian negative divergence. This divergence is small enough to not produce substantial effects in the compressor as the pulse is compressed to 120 fs FWHM, yet allows the pulse to have a waist at the center of the vapor source, nearly 40 meters downstream.

After exiting the compressor, which is under high vacuum to prevent nonlinear effects in the air during propagation, the ionizing laser pulse continues to an out of plane periscope consisting of mirrors MP1 and MP2 then on to the merging point and virtual line.

A 1 percent bleed from MP1 is split, then nearly all of this pulse energy is sent to a single shot SHG autocorrelator to verify the pulse length of the laser pulse. Since the pulse duration is ~120 fs FWHM, the group delay dispersion from travelling through MP1 gives only a percent error on the pulse duration.

The bleed of the mirror just before the autocorrelator is sent to a single lens imaging system to capture the pulse transverse profile just after MP1. This distribution can be used as a ‘near field’ distribution, as it is nearly 30 meters upstream from the waist.  

\subsection{Virtual Laser Line Description}

The ionizing laser pulse arrives from the laser room’s MP2 mirror to a combination periscope combined of the mirrors MP3 and MP4, which is an in-plane periscope. The main pulse continues to the MP5 mirror and is reflected to the vapor source.

Both MP4 and MP5 extract 1\% transmission pulses from the main ionizing laser pulse.

Following the MP4 bleed pulse, also known as the virtual line pulse: MP4 and a diagnostic mirror form an out of plane periscope that rotates the pulse and polarization. The virtual line pulse now continues upstream along a delay line to match its path length from MP4 to the entrance virtual line camera (VLC Entrance, named C1 on the sketch of the setup in the main paper) with the path length of the main ionizing pulse from MP4 to the entrance of the vapor source. Beamsplitters are placed in front of VLC Entrance and Center such that the energy of the incoming virtual line pulse is split evenly among the three cameras. The center virtual line camera (VLC Center, C2 on the sketch in the paper) and exit virtual line camera (VLC Exit, C3 on the sketch in the paper) are placed 5 and 10 meters after VLC Entrance, respectively, to match the paths of the main pulse through the vapor source. Neutral density filters protect the cameras from damage and saturation. There is one common filter just after the MP4 bleed, as well as independent filters on each of the virtual line cameras. 
 
The bleed from MP5 is sent via an out of plane periscope to an energy meter (denoted by $E_{in}$ on the sketch in the paper). This energy meter is calibrated with respect to the main ionizing pulse at low energy and then is the input pulse energy diagnostic.

\subsection{Laser Beam Dump Diagnostic Table}

The ionizing laser pulse exits the vapor source out of a 10 mm diameter aperture. The pulse then travels 10 meters to a wedge optic with a beam dump placed behind it. The polarization of the laser pulse is vertical and the wedge allows for approximately 0.5\% reflection from its front surface, while the rest of the pulse travels through the wedge optic and is dumped. This sampled pulse then travels downward to a second mirror that forms an in-plane periscope. The sampled pulse then continues onto a beamsplitter that makes it travel through a two lens relay type telescope with an f-number of 20 that is imaging the output of the vapor source. This image is relayed to the ``PickOff'' camera that samples the output transverse distribution of the laser spot at the exit of the vapor source. A filter wheel ensures that the camera image is not saturated.

Most of the sampled laser pulse travels on to an energy meter (denoted by $E_{out}$ on the sketch in the paper) to determine the transmitted pulse energy after travelling through the vapor source.

\section{Calibration of the energy measurements}

Because the input and output energy meters that measured $E_{in}$ and $E_{out}$ (type Gentec-EO QE8SP-B-MT) used a fraction of the laser pulse energy, a calibration process had to be executed before the experiments. The absolute calibration of the energy meter for $E_{in}$ was performed by placing an energy meter directly into the laser beam (when the vacuum system was opened) and recording the reading of $E_{in}$ together with that of the direct energy meter. 

The energy meter for $E_{out}$ was then calibrated against the input energy meter using the laser pulses propagating through the vapor source when it was cold and contained only an insignificant amount of residual vapor for the scales of interest. Using the noise equivalent energy of this energy meter head and the ratio of the pulse energy it received (this ratio having been determined by the calibration), we estimate the noise floor of the output energy measurement to be 0.7 mJ, its relative accuracy for energies well above this to be about $\pm$5\%.

The amount of residual rubidium vapor left in the vapor source is difficult to determine precisely. The vapor source itself is a dynamical system, when the valves of the two reservoirs are opened, vapor flows into the source and eventually leaves it at the two 10 mm diameter holes at the two ends of the tube \cite{Oz2014}. To estimate the vapor density that flows out of the reservoirs one can use the vapor-pressure model of \cite{Alcock1984}, and this is a good estimation of the vapor density  that should be in the vapor source at a given temperature {\em with the reservoir valves open}. At 25 \textdegree C this model yields a density of $\mathcal{N}_{res} = 1.27\cdot10^{10}\mathrm{~cm}^{-3}$ and this was used for residual vapor density even though {\em the reservoir valves were closed} for these measurements. The vapor density with the valves closed is mostly determined by the remnants of rubidium deposited on the walls, which depends on how the system has been cooled down and how long it has been kept at a given temperature. The true vapor density was certainly much smaller (possibly orders of magnitude smaller) than the value quoted above. However, the vapor density measurement utilized at the two ends of the vapor source was geared to be sensitive around the  $10^{14}-10^{15}\mathrm{~cm}^{-3}$ operational density, it had a noise floor of about  $10^{13}\mathrm{~cm}^{-3}$, much higher than the residual density.

For the purpose of calibration, the energy loss of the pulse propagating through residual vapor was estimated by assuming that it loses the energy of three photons for each atom in a cylinder of 1 mm radius and 10 m length, which, for the above density $\mathcal{N}_{res}$ amounts to about $3\cdot 10^{-7}$ J. Noting the overestimation of the true density, this is a very conservative estimate and is still three orders of magnitude smaller than the noise floor of the output energy meter. On the other hand, should we have a residual vapor density whose energy absorption is comparable to the noise floor of the measurement ($\mathcal{N}\approx10^{13}{\mathrm ~cm^{-3}}$), we would be able to detect (though not accurately measure) that with the vapor density measurement.


